\documentclass[12pt,preprint]{aastex}

\def\pp{\parshape 2 0.0truecm 16.25truecm 2.0truecm 14.25truecm}  
\newcommand{\be}{\begin{equation}}
\newcommand{\ee}{\end{equation}}
\newcommand{\xin}{{\xi_{1}}}
\newcommand{\xout}{{\xi_{2}}}
\newcommand{\xistar}{{\xi_\ast}} 
 
\newcommand{\thalf}{ {\widetilde \tau}_{1/2} }  
\newcommand{\evar}{{ \widetilde e}} 
\newcommand{\tpar}{{ {\hat t}_p }}

\newcommand{\xilmc}{{ \xi_{\rm lmc} }}
\newcommand{\rlmc}{{ r_{\rm lmc} }}
\newcommand{\massin}{ M_{\rm in} }  
\newcommand\lta{\,\raise 0.3 ex\hbox{$ < $}\kern -0.75 em
 \lower 0.7 ex\hbox{$\sim$}\,}
\newcommand\gta{\,\raise 0.3 ex\hbox{$ > $}\kern -0.75 em
 \lower 0.7 ex\hbox{$\sim$}\,}  

\begin{document} 

\title{ORBITS IN EXTENDED MASS DISTRIBUTIONS: \\
GENERAL RESULTS AND THE SPIROGRAPHIC APPROXIMATION} 
 
\author{Fred C. Adams$^{1,2}$ and Anthony M. Bloch$^{1,3}$ } 
 
\affil{$^1$Michigan Center for Theoretical Physics \\ 
Physics Department, University of Michigan, Ann Arbor, MI 48109} 

\affil{$^2$Astronomy Department, University of Michigan, Ann Arbor, MI 48109} 

\affil{$^3$Mathematics Department, University of Michigan, Ann Arbor, MI 48109} 

\begin{abstract} 

This paper explores orbits in extended mass distributions and develops
an analytic approximation scheme based on epicycloids (spirograph
patterns). We focus on the Hernquist potential $\psi = 1/(1 + \xi)$,
which provides a good model for many astrophysical systems, including
elliptical galaxies (with an $R^{1/4}$ law), dark matter halos (where
N-body simulations indicate a nearly universal density profile), and
young embedded star clusters (with gas density $\rho \sim \xi^{-1}$).
For a given potential, one can readily calculate orbital solutions as
a function of energy and angular momentum using numerical methods. In
contrast, this paper presents a number of analytic results for the
Hernquist potential and proves a series of general constraints showing
that orbits have similar properties for any extended mass distribution
(including, e.g., the NFW profile). We discuss circular orbits, radial
orbits, zero energy orbits, different definitions of eccentricity,
analogs of Kepler's law, the definition of orbital elements, and the
relation of these orbits to spirograph patterns (epicycloids). Over a
large portion of parameter space, the orbits can be adequately
described (with accuracy better than 10\%) using the parametric
equations of epicycloids, thereby providing an analytic description of
the orbits. As an application of this formal development, we find a
solution for the orbit of the Large Magellanic Cloud in the potential
of our Galaxy. 

\end{abstract}

\keywords{stellar dynamics, celestial mechanics --- methods:
analytical --- galaxies: kinematics and dynamics, halos ---
stars: formation}

\section{INTRODUCTION} 

Orbits are a fundamental component of dynamical astrophysical systems,
including dark matter halos, galaxies, star clusters, and solar
systems. For a single point mass, the source term for a Keplerian
potential, orbits are well-known conic sections and have been the
cornerstone of solar system dynamics for centuries (starting with
Newton 1687; for a modern treatment, see Murray \& Dermott 1999).
Many astrophysical systems have extended mass distributions, which
give rise to more complex potentials and more complicated orbits
(Binney \& Tremaine 1987).  Although an enormous amount of work has
been done on understanding the dynamics of such extended mass
distributions, the orbits themselves are not as well studied. In this
contribution, we calculate the orbits for a class of extended mass
distributions that provide good working models for dark matter halos,
elliptical galaxies, and embedded young star clusters, albeit in
rather different regimes of parameter space (see also van den Bosch et
al. 1999 for a consideration of orbits in an isothermal potential).
Along with our improved characterization of the orbits, we develop an
effective analytic approximation scheme that describes the orbits in
terms of epicycloids (spirograph patterns) and thereby obtain a set of
parametric equations for the orbits. In addition to making dynamical
calculations easier, this analytic approximation scheme adds to our
conceptual understanding of the orbits --- in much the same way that
knowing Keplerian orbits are conic sections provides increased 
perception. 

In this paper, we focus on the Hernquist potential, which arises from
a density profile of the form 
\be 
\rho = {\rho_0 \over \xi (1 + \xi)^3} \, ,  
\label{eq:rhobasic} 
\ee 
where the dimensionless radius $\xi = r/r_s$ and $r_s$ is the length
scale of the potential. This density profile has a corresponding
analytic form for its distribution function (Hernquist 1990). One goal
of this paper is to characterize the orbits for bodies traveling
within this extended density distribution. The motivation for this
effort is that particular subclasses of these orbits describe the
motion of bodies in many astrophysical systems.  The original
application of this potential (Hernquist 1990) was to approximate the
$R^{1/4}$ law for elliptical galaxies (de Vaucouleurs 1948).  In
addition, density profiles of this form appear in dark matter halos of
galaxies and galaxy clusters, and in the molecular cloud cores that
form stellar groups and clusters.  Both of these latter applications
require some explanation:

Over the past decade (starting with Navarro, Frenk, \& White 1997;
hereafter NFW), numerical simulations of cosmic structure formation
have shown that the density distributions of dark matter halos
approach a nearly universal form (see also Crone et al. 1996, Moore et
al. 1998, Bullock et al. 2001).  Although the original form for this
universal profile (NFW) is not as steep as that of equation
(\ref{eq:rhobasic}) at large radii, subsequent work indicates that the
Hernquist density profile provides a good description for the
asymptotic structure of dark matter halos (Busha et al. 2003, 2004).
One advantage of using equation (\ref{eq:rhobasic}) is that the total
mass of the system is finite (whereas the integral of the original NFW
density profile diverges). One long term goal of structure formation
studies is to understand the dynamics of these halo systems, including
both the origin of the universal profile and its consequences. Toward
that end, we must describe the orbital motions of test particles 
moving in the gravitational potential created by the universal density
profile (where the ``test particles'' can range from individual dark 
matter particles with mass $m_W \sim 100$ GeV to small satellite
galaxies with mass $M_{sat} \sim 10^9$ $M_\odot$).

Molecular cloud cores that form stellar clusters provide another
application. They are inferred to have equations of state softer than
isothermal (e.g., Larson 1985; Jijina et al. 1999) and hence density
profiles of the approximate form $\rho \sim 1/r$, which coincides with
the Hernquist density profile at small radii. The orbits found for the
inner limit of the density profile (\ref{eq:rhobasic}) are thus
applicable to star forming regions. Most stars form in groups/clusters
(Lada \& Lada 2003, Porras et al. 2003) and subsequent stellar orbital
motions determine the degree to which the environment affects star and
planet formation (e.g., Adams \& Myers 2001; Adams \& Laughlin 2001).

Since the potentials considered here are spherically symmetric, motion
is confined to a plane (in the absence of perturbations) and the
orbits can be calculated to high accuracy using numerical methods.
This paper takes a complementary approach and determines the nature of
the orbits using analytic methods as far as possible and develops an
analytic approximation scheme based on spirograph patterns.  Our goal
is to understand the orbits at the same level that we now understand
Keplerian orbits, in spite of the greater mathematical complexity.
Several subtleties arise that cannot be captured through numerical
solutions alone: The orbital eccentricity can be defined in several
ways, and each definition agrees with the Keplerian result in
different aspects. The orbits do not close and their turning angles
satisfy $\Delta \theta < \pi$ (for a half orbit) and this angular
deficit remains even in the circular limit. The maximum angular
momentum, the maximum turning angle, and the radius of the circular
orbit can all be determined analytically as a function of energy.
These results allow for many properties of these orbits to be found
analytically and add to our understanding of how the orbits depend on
the underlying potential (see \S 2).

Although a large number of results can be obtained analytically, a
complete description of the orbital shapes requires numerical
evaluation. Nonetheless, the orbit shapes are similar to the familiar
form of epicycloids, often known as spirograph patterns (and not to be
confused with epicycles -- see below). This paper takes this analogy
one step further (\S 3) and shows that for a large fraction of the
relevant parameter space, the orbits in the Hernquist potential can be
modeled with high accuracy (better than 10\%) using the parametric
equations of epicycloids. The resulting spirographic approximation
provides an analytic description of the orbital shape.  This analytic
description, in turn, allows for other orbital quantities to be
calculated analytically. For example, we show the effects of geometry
on orbital decay by integrating $\Delta j/j$ over an orbit (\S 3.3)
and we use the spirographic approximation to model the orbit of the
Large Magellanic Cloud (LMC) in the potential of the Galaxy (\S 4).

In addition to the applications presented in this paper, the analytic
approximations developed here can be used in a wide variety of other
contexts. For example, an analytic description of the orbits can be
useful in galactic stability theory (e.g., Evans \& Read 1998), where
the perturbed distribution function is given as an integral over the
equilibrium orbit. Another application of analytic orbits is to find
steady-state distribution functions, which depend on the isolating
integrals of motion (or the constants in the spirographic
approximation developed here). This analytic description of orbits
will also be useful in assessing the effects of cluster environments
on the star formation process: In this context, the orbits of newly
formed stars determine how much radiation impinges on their
circumstellar disks and their probability of experiencing scattering
interactions; this physical processes, in turn, determine how the
cluster environment can compromise planet formation.

Although extended mass distributions do not always have the form of
equation (\ref{eq:rhobasic}), we argue that the orbits found here
provide good baseline models for orbital behavior in more general
potentials. In particular, generic orbits in extended mass
distributions will be more like those of the Hernquist model than the
conic sections resulting from a Keplerian potential. Further, such
generic orbits can be described through the spirographic approximation
developed in this paper. Orbits in general extended mass distributions
share a number of traits: For mass distributions of physical interest,
the topology of the manifold of orbital solutions is simple (the same
as the $S_1 \times S_1$ structure of the Kepler problem; see Smale
1970ab), although the geometry of the orbits is more complicated. One
defining characteristic of these orbits -- turning angles for a half
orbit bounded by $\pi/2 < \Delta \theta < \pi$ -- holds for general
extended mass distributions (Contopoulos 1954).  This paper proves
additional results concerning generic orbits in extended mass
distributions (\S 5). We show that the eccentricity can be defined in
multiple ways, with the definitions coincident only for a Keplerian
potential. We also show that orbital shapes in generalized extended
mass distributions are confined to be relatively close to those of
epicycloids (spirograph patterns). We validate this claim by finding a
set of bounds on orbital shapes that apply to all physical potentials,
and by explicit construction of orbits -- and comparison with the
spirographic approximation -- for a collection of well known
potentials.

\newpage 

\section{ORBIT SOLUTIONS} 

In this section we calculate orbits for bodies moving in the potential
of the Hernquist profile (with the density distribution of eq. 
[\ref{eq:rhobasic}]). We define the total depth of the gravitational
potential well $\Psi_0 \equiv 2 \pi G \rho_0 r_s^2$ and the total mass
$M_\infty \equiv 2 \pi r_s^3 \rho_0$, so that the corresponding mass
profile, force profile, and gravitational potential take the forms 
\be 
M(\xi) = M_\infty {\xi^2 \over (1 + \xi)^2} \, , \qquad 
F(\xi) = {\Psi_0/ r_s \over (1 + \xi)^2} \, , \qquad 
\Psi(\xi) = {\Psi_0 \over 1 + \xi} \, . 
\label{eq:profiles} 
\ee 
All quantities are taken to be positive (with the proper signs 
inserted as necessary). 

\subsection{General Orbits} 

The basic formulation for orbits in spherical mass distributions is
well known (Binney \& Tremaine 1987); this section defines notation
and presents analytic results for orbits in the Hernquist potential.
An orbit in any potential $V(r)$ can be determined from the
differential equation 
\be
{d \theta \over dr} = {1 \over r} 
\Bigl[ 2 (E - V) r^2 / j^2 - 1 \Bigr]^{-1/2} \, , 
\ee
where $E$ is the energy and $j$ is the specific angular momentum.
This paper focuses on bound orbits with negative energy. We define 
dimensionless variables for the energy ($\epsilon$) and angular 
momentum ($q$) according to 
\be
\epsilon \equiv |E| / \Psi_0 \qquad {\rm and} \qquad 
q \equiv j^2 / 2 \Psi_0 r_s^2 \, . 
\ee  
The differential form of the orbit equation can then be written 
$\xi d\theta/d\xi = [q / f(\xi) ]^{1/2}$, so that much of the 
dynamics is determined by the properties of the cubic equation 
\be 
f(\xi) = - \epsilon \xi^2 + {\xi^2 \over 1 + \xi} - q \, . 
\label{eq:cubic} 
\ee 
For a given energy $\epsilon$ and angular momentum parameter $q$, the
orbit exists within the range of $\xi$ for which $f(\xi)$ is positive. 
The two positive zeroes of $f(\xi)$ thus correspond to the radial
turning points of the orbit and are denoted here as $\xi_1$ and
$\xi_2$. The third zero of the cubic is always negative for the
physical range of parameter space and we denote this root as $a \equiv
- \xi_0 > 0$. Finding the roots of the equation (\ref{eq:cubic}) is
straightforward, but cumbersome (e.g., Beyer 1980). However, the
inverse relations can be found simply: 
$$\epsilon = {\xi_1 + \xi_2 + \xi_1 \xi_2 \over 
(\xi_1 + \xi_2) (1 + \xi_1 + \xi_2 + \xi_1 \xi_2) } \, , \qquad 
q = { (\xi_1 \xi_2)^2 \over 
(\xi_1 + \xi_2) (1 + \xi_1 + \xi_2 + \xi_1 \xi_2) } \, , $$
\be 
{\rm and} \qquad 
a = - \xi_0 = { \xi_1 \xi_2 \over \xi_1 + \xi_2 + \xi_1 \xi_2 } \, . 
\label{eq:inverseroots} 
\ee 

One quantity of interest for a given orbit is the angle $\Delta
\theta$ through which the orbit turns over one half cycle, e.g., as 
the orbit moves from $\xi_1$ to $\xi_2$. This angle is given by 
\be 
\Delta \theta = \sqrt{q} \, \int_{\xi_1}^{\xi_2} {d\xi \over \xi} 
\bigl[ f(\xi) \bigr]^{-1/2}  \, .  
\label{eq:orbang} 
\ee  
A related quantity of interest is the orbital half period, denoted
here as $\tau_{1/2}$.  The natural time scale of this potential is
$\tau_0 \equiv$ $r_s (2 \Psi_0)^{-1/2}$.  The dimensionless time scale
$\thalf \equiv \tau_{1/2}/\tau_0$ can be written in integral form 
\be 
\thalf = \int_{\xi_1}^{\xi_2} \, \xi \, d\xi \, 
\bigl[ f(\xi) \bigr]^{-1/2}  \, .   
\label{eq:orbtime}
\ee

For a given dimensionless energy $\epsilon$, orbits can have a maximum
specific angular momentum and hence a maximum value of $q$. This
maximum value takes the form 
\be 
q_{max} = {1 \over 8 \epsilon} 
{ (1 + \sqrt{1 + 8 \epsilon} - 4 \epsilon )^3 \over 
(1 + \sqrt{1 + 8 \epsilon})^2 } \, . 
\label{eq:qmax} 
\ee 
In the limit of low energy $\epsilon \to 0$, we recover the Keplerian 
result $4 \epsilon q_{max}$ = 1.  In the opposite limit of $\epsilon
\to 1$, we write $\epsilon$ = $1 - \delta$ and find the limiting form
$q_{max} = 4 \delta^3/27$.

For a given energy $\epsilon$, the minimum of the cubic function
$f(\xi)$ is independent of the angular momentum variable $q$. The
location of this minimum, denoted here by $\xistar$ where $df/d\xi
(\xistar)$ = 0, is given by 
\be
\xistar = {1 - 4 \epsilon + \sqrt{1 + 8 \epsilon} \over 4 \epsilon} \, .  
\label{eq:xistar} 
\ee
In the limit of low energy $\epsilon \to 0$, we recover the Keplerian
form $\xistar = 1/2 \epsilon$. As a result, we can identify $\xistar$
as playing the same role that the semi-major axis plays for bound
orbits in Keplerian potentials (although these orbits are not
ellipses).  Notice also that $\xi_\ast$ is the radius of the circular
orbit for a given energy $\epsilon$.

For given values of energy and angular momentum ($\epsilon$, $q$), the
orbit takes a specified form. One such orbit is shown in Figure 1 for
the case $\epsilon$ = 0.10 and $q = q_{max}/2$.  For each half
oscillation in the radial direction, as $\xi$ ranges from $\xout$ down
to $\xin$, the orbit sweeps through an angle $\Delta \theta$, which
lies in the range $\pi/2 \le \Delta \theta \le \pi$. In the Keplerian
limit ($\epsilon \to 0$), $\Delta \theta \to \pi$. In the limit of
radial orbits ($q \to 0$), the angle $\Delta \theta \to \pi/2$.
Between these limits, the angular displacement takes on intermediate
values that depend on energy and angular momentum.  These results are
illustrated in Figure 2, which shows $\Delta \theta$ as a function of
$q/q_{max}$ for a collection of energy values. One curious result is
that the angle $\Delta \theta$ does not approach $\pi$ in the limit of
circular orbits (see eq. [\ref{eq:thetalim}] below).  For the set of
curves shown in Figure 2, specifically for energy and angular momentum
in the range $0.2 \le \epsilon \le 1$ and $10^{-6} \le q/q_{max} \le 1$, 
the turning angle $\Delta \theta$ can be approximated with a fitting 
function of the form 
\be 
{\Delta \theta \over \pi} = {1 \over 2} + 
\big[ (1 + 8 \epsilon)^{-1/4} - {1 \over 2} \bigr] 
\bigl[ 1 + { \ln (q/q_{max}) \over 6 \ln 10} \bigr]^{3.6} \, . 
\label{eq:fitting} 
\ee 
This function provides an approximation to the turning angle with 
an accuracy of about $2.5\%$ over the previously stated portion of 
parameter space.

Figure 3 shows the orbital half period, i.e., the time required for
the orbiting body to travel from the inner turning point to the outer
turning point. In the figure, the half period is normalized by the
factor $\epsilon^{3/2}$, which represents the variation expected in
the Keplerian limit, and by the factor $[\cos^{-1} \sqrt{\epsilon} -
\sqrt{\epsilon} (1 - \epsilon)^{1/2} ]$ that gives the correct orbit
time in the limit of zero angular momentum. The resulting (normalized)
half period is nearly independent of angular momentum, and nearly the
same as that of the radial orbit. The orbiting body spends most of its
time at large $\xi$, where $q$ has little effect, so the result is
almost independent of $q$. The half period does, however, depend on
the energy $\epsilon$.

Given an energy and angular momentum $(\epsilon, q)$, and hence the
radial turning points, one can define a generalized ``eccentricity''
in terms of $\xin$ and $\xout$. Unlike the Keplerian case, the
definition of eccentricity leads to some ambiguities. Two reasonable
definitions of eccentricity are 
\be 
e \equiv {\xout - \xin \over 2 \xistar} \,  
\qquad {\rm and} \qquad 
\evar \equiv {\xout - \xin \over \xout + \xin} \,  \, . 
\label{eq:eccent} 
\ee  
At low energies $\epsilon \to 0$, the orbits become Keplerian and both
definitions of the eccentricity coincide with the traditional one.  
In general, however, the two definitions do not coincide (see \S 5.3).  
The first definition of eccentricity $e$ has the advantage in that it
includes the radius $\xistar$ of the circular orbit and the turning
points represent symmetric departures from $\xistar$ (analogous to the
Keplerian case). In contrast, the second definition $\evar$ does not
correctly describe small departures about the circular orbit because
$\xout + \xin \ne 2 \xistar$. On the other hand, the second definition
allows for the full range of eccentricities $0 \le \evar \le 1$,
whereas the first definition has a maximum eccentricity so that $0 \le
e \le e_{max}$ (for all nonzero energies $\epsilon > 0$). This maximum
eccentricity occurs for radial orbits ($q \to 0$), where the outer turning
point $\xout \to (1-\epsilon) /\epsilon$ and the inner turning point
$\xin \to 0$, and is given by 
\be 
e_{max} = 
{2 (1 - \epsilon) \over 1 - 4 \epsilon + \sqrt{1 + 8 \epsilon} } \, . 
\ee 
In the Keplerian limit $\epsilon \to 0$, one recovers the expected 
result that $e_{max} \to 1$. In the inner limit $\epsilon \to 1$, 
one finds $e_{max} \to 3/4$. In between, the maximum eccentricity 
is a continuous monotonic function of dimensionless energy $\epsilon$.   

\subsection{Limiting Forms of the Solutions} 

For the cases of circular orbits, radial orbits, and zero energy
orbits, the solutions reduce to limiting forms that can be readily
evaluated.  For circular orbits, one can solve for the relationship
between the orbital period $\tau$ and radius. The resulting ``Kepler
Law'' for circular orbits takes the form 
\be 
\tau^2 = {2 \pi \over G \rho_0} \xi (1 + \xi)^2 \, . 
\ee 
For a circular orbit at radius $\xi$, the energy and angular momentum
are given by $\epsilon$ = $(1 + \xi/2) (1+\xi)^{-2}$ and $q = (1/2)
\xi^3 (1+\xi)^{-2}$.  As a result, we can write the orbital period in
terms of the dimensionless energy so that 
\be 
\tau^2 = {\pi \over G \rho_0} \, {1 \over \epsilon} \, 
\Bigl[ {1 + 4 \epsilon + \sqrt{1 + 8 \epsilon} \over 8 \epsilon^2} 
- 1 \Bigr] \, . 
\ee  
As defined here, the period $\tau$ is the time required for the orbit
to make a complete circuit, i.e., for the angle $\theta$ to trace
through $2 \pi$ radians. However, the angular integral
(eq. [\ref{eq:orbang}]) does {\it not} approach $\pi$ in the limit of
circular orbits (as it does for Keplerian orbits). The angular 
displacement is less than $\pi$ and depends on energy according to 
\be 
\lim_{q \to q_{max}} \Delta \theta = \pi (1 + 8 \epsilon)^{-1/4} \, . 
\label{eq:thetalim} 
\ee
In the Keplerian limit $\epsilon \to 0$, $\Delta \theta \to \pi$, as
expected. For $\epsilon \ne 0$, the formula (\ref{eq:thetalim}) agrees
with the limit points of the curves shown in Figure 2. Since circular
orbits are stable in this potential, small perturbations about the
circular orbit lead to oscillations at the epicyclic frequency
$\kappa$, which takes the form $\kappa^2$ = $2 \pi G \rho_0 (3 + \xi) 
\xi^{-1} (1 + \xi)^{-3}$. 

Another class of orbits that can be considered separately are those
with zero angular momentum, i.e., radial orbits. In this case, the 
angular coordinate is no longer of interest and the orbits are 
characterized by their energy $\epsilon$. Each energy has a
corresponding infall time scale $\tau$, defined to be the time
required to fall from the outer turning point $\xi_2$ to the center. 
This time scale can be evaluated to obtain the result 
\be 
\tau = (4 \pi G \rho_0)^{-1/2} \, 
\epsilon^{-3/2} \Bigl[ \cos^{-1} \sqrt{\epsilon} 
+ \sqrt{\epsilon} \sqrt{1 - \epsilon} \Bigr] \, . 
\label{eq:analtau} 
\ee 
For radial orbits, the position along the orbit is specified by the
radial coordinate so the angle $\theta$ is no longer necessary. 
Nonetheless, the angle of the orbit remains defined and the half 
angle of the orbit (eq. \ref{eq:orbang}) reaches a definite limit. 
As $q \to 0$, we find $\Delta \theta \to \pi/2$.  

Finally, we consider zero energy orbits. For Keplerian potentials,
such orbits are parabolas. Here, the corresponding orbits are more
complicated, but simple solutions can be found for limiting cases.
The key parameter for a zero energy orbit is its inner turning point,
which is given by 
\be 
\xi_1 = {q + \sqrt{q^2 + 4 q} \over 2} \, . 
\ee  
One limiting form of this solution occurs when $\xi \ll 1$ for the
entire orbit. This solution would also result for the case in which a
density profile of the form $\rho \sim 1/r$ extends out to infinity.
In this limit, the orbit equation can be solved to obtain $\sin \theta
= \sqrt{q}/ \xi$, i.e., the orbital path is a straight line defined by
$y = \xi \sin \theta = \sqrt{q}$ = {\sl constant}.  For orbits that
remain in the outer halo where $\xi \gg 1$, we can also simplify the
expression to obtain 
\be 
\cos \Bigl[ \bigl( {q+1 \over q} \bigr)^{1/2} \theta \Bigr] = 
1 - {2 (q+1) \over \xi } \, . 
\ee   

\subsection{Orbital Elements} 

A particle in orbit can be described by six phase space variables,
which are usually taken to be the position vector ${\bf x} = (x,y,z)$
and the velocity vector ${\bf v} = (v_x, v_y, v_z)$. For the Kepler
problem, one can replace these phase space variables with osculating
orbital elements, i.e., the elliptical (Keplerian) orbit that has the
same instantaneous position and velocity vectors (Murray \& Dermott
1999). For trajectories in extended mass distributions, the phase
space variables can be written in terms of the elements of the orbits
found here.  Since the potential is spherically symmetric and angular
momentum is conserved, orbital motion is confined to a plane. The two
angles that define the unit normal of the orbital plane $(\theta_p,
\phi_p)$ thus provide the first two orbital elements.  The rest of
this discussion is confined to this orbital plane.  Although the
assignment of the remaining orbital elements is not unique, we present
a sensible prescription in what follows.

Within the plane, the orbital shape is determined by the energy
$\epsilon$ and angular momentum, specified here by $q$. The pair
$(\epsilon,q)$ thus can be used as the next two orbital elements.
However, other choices are available: The energy $\epsilon$ can be
replaced with the effective semi-major axis $\xistar$, as defined by
equation (\ref{eq:xistar}). For a given energy and angular momentum,
the cubic equation (\ref{eq:cubic}) defines the turning points $\xi_1$
and $\xi_2$ of the orbit, so that the generalized eccentricity (eq.
[\ref{eq:eccent}]) is well defined and can be used as the other
variable. One can thus replace the variables $(\epsilon,q)$ with the
pair $(\xistar,e)$, which provides the closest analogy to the Kepler
problem. Alternately, one can use the turning points themselves,
$(\xi_1, \xi_2)$, to specify the orbital shape. Recall that the cubic
equation that defines the turning points has three real roots in the
regime of interest; because the problem has only two parameters
($\epsilon, q$), specification of the turning points ($\xi_1, \xi_2$)
is sufficient to define the equation, and the third root can be
written in terms of the turning points (eq. [\ref{eq:inverseroots}]).

The two remaining orbital elements must specify the orientation of the
orbit within the plane and the position of the particle along its
orbit. In the custom of solar system dynamics, the term longitude is
used to denote an angle that is measured relative to a fixed location
in inertial space. Here we can specify the orientation of the orbit in
terms of the longitude $\varpi$ of the inner turning point $\xi_1$
(the longitude of pericenter).  Like the elliptical orbits in the
Kepler problem, the longitude of pericenter can precess.  Unlike the
case of the Kepler problem, orbits in extended mass distributions are
not closed and the longitude $\varpi$ of pericenter changes with every
orbit. A full orbit turns through an angle $2 \Delta \theta < 2 \pi$, 
so the pericenter precesses backwards at the rate of $2 (\pi - \Delta 
\theta)$ per orbit. The longitude of pericenter is given by $\varpi =
2(\pi - \Delta \theta) n + \varpi_0$, where $\varpi_0$ is the starting
value ($t=0$) and $n$ is the number of full orbits that have elapsed.
Notice that we can use either the starting longitude $\varpi_0$ or the
current longitude of pericenter $\varpi$ as the next orbital element.
For completeness, we can also define a continuous longitude of
pericenter through the relation $\varpi_{\rm c} \equiv 2(\pi - \Delta
\theta) t/\tau + \varpi_0$, where $\tau$ is the (full) orbital period.
Finally, the location of the particle along the orbit is given by its
angular displacement $\theta$.  Again following the custom of solar
system dynamics, the angular displacement variable can be called the
{\it anomaly}. The angle $\theta$ is thus the true anomaly and can be
measured in two different ways. The natural definition is to measure
$\theta$ from the inner turning point (or the outer turning point) for
the current longitude of pericenter $\varpi$. However, one could also
measure the angle with respect to the starting longitude of pericenter
$\varpi_0$ , in which case the anomaly is denoted as $\theta_0$. We
can also define a mean anomaly $M = \omega (\Delta t)$, where $\Delta
t$ is the time since pericenter passage and the mean motion variable
$\omega \equiv (\Delta \theta)/\tau_{1/2}$ (see eqs. 
[\ref{eq:orbang}--\ref{eq:orbtime}]).

A collection of possible orbital elements is summarized in Table 1.
Here we assume that the orbital plane is specified so that only four
phase space variables remain. The usual variables $[(x,y), (v_x,
v_y)]$ can be replaced with various choices for the osculating orbital
elements listed in Table 1. For each pair of shape parameters, any of
the possible pairs for the orientation and anomaly are viable choices. 
The last line of the table lists the variables required to describe
the orbit as an epicycloid (spirograph pattern), as discussed in the
following section. 

\bigskip 

\centerline{\bf Table 1: Orbital Elements}  
\medskip 

\begin{center}
\begin{tabular}{ccc}
\hline 
\hline
Shape Parameters & (Orientation, Anomaly) \\ 
\hline 
($\epsilon$, $q$) & ($\varpi_0, \theta_0$) \\ 
($\xistar$, $e$) & ($\varpi, \theta$) \\ 
($\xi_1$, $\xi_2$) & ($\varpi_{\rm c}, M$) \\ 
\hline 
($\alpha, \beta, \gamma$) & ($\phi, \tpar$) \\ 
\hline 
\hline 
\end{tabular}
\end{center}   

\section{EQUIVALENT SPIROGRAPHIC ORBITS}  

The orbits found here for the Hernquist potential look much like
epicycloids, more generally known as spirograph patterns. In this
section, we explore the relationship between the actual orbits and the
closest equivalent spirographic orbits. We find a quantitative measure
of the difference between the physical orbits and the spirographic
approximation, and show that the spirographic treatment is valid over
a large portion of parameter space. This finding allows for a class of
physical orbits to be described using simple parametric -- and hence
analytic -- equations. The resulting analytic description facilitates
a greater physical understanding of the orbital dynamics. For example,
the effects of dissipation (friction) on the orbits can be modeled
analytically using the spirographic approximation (see \S 3.3).

The epicycloids discussed here are not the same as epicycles, which
are often considered in connection with astronomical orbits. In the
original Ptolemaic definition, an epicycle is the curve generated by a
small circle that rotates while it moves along the circumference of a
larger circle. Epicycloids are curves generated by a smaller circle
rotating inside (or outside) a larger circle, with a drawing point
located a given distance away from the center of the smaller circle.
In modern astronomy, the epicyclic frequency is the frequency of
radial oscillation for a body in a circular orbit, when perturbed in
the radial direction by an infinitesimal amplitude [so that $\kappa^2
= r^{-3} d(r^4\Omega^2)/dr$]. As a result, the inclusion of epicyclic
motion (radial oscillations with frequency $\kappa$) provides a first
order correction to a circular orbit in an arbitrary potential [which
determines the rotation curve $\Omega(r)$]. Many previous studies have
explored the epicyclic approximation and the inclusion of higher order
terms (e.g., Shu 1969, Kalnajs 1979, Dehnen 1999). In contrast, this
present work explores epicycloids as approximations to physical orbits.

\subsection{Parametric Description of the Orbits} 

In a fixed plane, here the orbital plane, the parametric equations for
an epicycloid (a generalized spirograph pattern) can be written in the
form 
$$ 
x(\tpar) = (\alpha - \beta) \cos \tpar +  
\gamma \cos [(\alpha - \beta) \tpar /\beta] \, , 
$$ 
\be 
y(\tpar) = - (\alpha - \beta) \sin \tpar + 
\gamma \sin [(\alpha - \beta) \tpar /\beta] \, , 
\label{eq:spirodef} 
\ee  
where $\alpha$ is the radius of the larger spirograph circle, $\beta$
is the radius of the smaller circle, and $\gamma$ is the distance from
the center of the smaller circle to the drawing point (e.g., see the
website of Little 1997,\footnote{Little, D. P. 1997,
http://www.math.dartmouth.edu/dlittle/java/SpiroGraph/} which includes
interactive drawing capability).  These length scales should be
considered dimensionless; they can be converted into physical units by
specifying a length scale ($r_s$).  Keep in mind that the variable
$\tpar$ is a parameter and does not represent physical time (its
relationship to time is elucidated below).  With this form of the
parametric equations, the spirograph pattern begins at $\tpar$ = 0
with its maximum radius with an angle $\theta = \arctan(y/x)$ = 0; the
pattern then turns (counterclockwise) through an angle $(1 -
\beta/\alpha) \pi$ as the radius decreases to its minimum value. The
parametric variable $\tpar$ ranges from 0 to $\beta \pi/\alpha$ over
this half orbit. Notice that the starting orientation of the orbit can
be changed by including a starting phase angle $\phi$ (equivalent to
an offset in the parametric time variable).

The parametric equations (\ref{eq:spirodef}) allow for two different
(equivalent) representations of physical orbits (see below). In this
treatment, we use the representation where $\gamma$ plays the role of
the semi-major axis. In this case, the transformation between the
spirographic variables and the physical variables -- and the inverse
transformation -- take the form 
$$
\xi_1 = \gamma - (\alpha - \beta) \, , \qquad 
\xi_2 = \gamma + (\alpha - \beta) \, , \qquad 
\Delta \theta = (\alpha - \beta) \pi / \alpha\, , 
$$ 
\be 
\gamma = {1 \over 2} (\xi_1 + \xi_2) , \qquad 
\alpha = {\pi \over 2 \Delta \theta} (\xi_2 - \xi_1) , \qquad 
\beta = {1 \over 2} [\pi/(\Delta \theta) - 1] (\xi_2 - \xi_1) \, . 
\label{eq:relate} 
\ee 
Since the turning points $\xi_1$ and $\xi_2$ of the physical problem
are related to the energy $\epsilon$ and angular momentum (specified
by $q$) through equation (\ref{eq:cubic}), the above relations
(eq. [\ref{eq:relate}]) implicitly define the spirograph parameters 
in terms of the energy and angular momentum.  The physical problem 
contains only two variables, which can be taken to be the energy and 
angular momentum ($\epsilon$, $q$) or the turning points ($\xi_1$,
$\xi_2$). The equivalent epicycloid is written in terms of three
parameters ($\alpha$, $\beta$, $\gamma$), but only two are
independent. Any such triple will define an epicycloid, but only those
related through equations (\ref{eq:relate}) are physically relevant.
The spirographic parameters are related to the energy and angular
momentum of the orbit via 
\be 
\epsilon = {\gamma^2 + 2 \gamma - (\alpha - \beta)^2 \over 
2 \gamma \bigl[ (1+\gamma)^2 - (\alpha - \beta)^2 \bigr] } 
\qquad {\rm and} \qquad 
q = {\bigl[ \gamma^2  - (\alpha - \beta)^2 \bigr]^2 \over 
2 \gamma \bigl[ (1+\gamma)^2 - (\alpha - \beta)^2 \bigr] } \, . 
\label{eq:eqspiro} 
\ee 

Notice that since $\xi_1, \xi_2 = \gamma \pm (\alpha - \beta)$, the
parameter $\gamma$ is akin to the semi-major axis of the orbit
(although this analogy is not exact because the orbits are not
ellipses). A natural definition of the eccentricity $e_s$ for a
spirographic orbit has the form 
\be 
e_s = {\alpha - \beta \over \gamma} \, . 
\label{eq:eccentspiro} 
\ee
Since $\gamma \ge (\alpha - \beta)$ for physically relevant orbits, 
the eccentricity $e_s \le 1$. 

In the Keplerian limit, the turning angle $\Delta \theta \to \pi$.
Thus, the Kepler limit corresponds to the limit in which $\beta \ll
\alpha$ and the turning points are given by $\xi_{1,2}$ = $\gamma \pm
\alpha$. In the limit of circular orbits, the above expressions
(eqs. [\ref{eq:eqspiro} and \ref{eq:eccentspiro}]) indicate that
$(\alpha - \beta) \to 0$. However, in order for the parametric
equations (\ref{eq:spirodef}) to trace a circle, the condition
$(\alpha - \beta)/\beta \to 1$ ($\alpha/\beta \to 2$) must also hold.
In order for both of these limits to apply, $\alpha, \beta \to 0$.
The limit of radial orbits is realized when the angular momentum
vanishes, which requires $\gamma=(\alpha - \beta)$, and $\Delta \theta
\to \pi/2$, which requires $\alpha = 2 \beta$; as a result, radial
orbits are given by $(\alpha,\beta,\gamma)$ = $(2x,x,x)$ for any value
of $x$.

The range of possible epicycloid patterns is much larger than the
range of orbital shapes that represent physical motion (in the
potential of an extended mass distribution). The allowed parameter
space is depicted in Figure 4 (where the parameter $\gamma$ is set
equal to unity).  For physical orbits, the turning angle for a half
orbit falls in the range $\pi/2 \le \Delta \theta \le \pi$; for our
chosen representation, this constraint restricts the parameter space
to the region $\beta < \alpha / 2$. Since the inner turning point must
be greater than zero, $\gamma > (\alpha - \beta)$. The allowed region
of the $\alpha-\beta$ plane is thus bounded from above by the dashed
line and from below by the dotted line. Above the dashed line where
$\beta = \alpha/2$, the turning angle is too small and the resulting
spirograph patterns are more open than those realized in physical
potentials. Below the dotted line, the inner turning point is formally
negative, so that the pattern turns ``in front of'' the center, rather
than orbiting around it. The resulting pattern is tighter than those
realized in physical orbits. The inset diagrams show the basic orbital
shapes for the different regions of parameter space. Notice that for
this formulation, $\gamma > \alpha, \beta$, and the drawing point of
these epicycloids extends beyond the smaller circle (so they cannot be
drawn with conventional spirographic wheels).

The region of the $\alpha-\beta$ plane above the dashed line and below
the dotted line (the upper right portion of Figure 4) corresponds to
the second representation of the orbits. In this region, the
complementary epicycloid pattern has the same shape, turning points,
and turning angle as the ``original'' pattern under the transformation 
\be 
\gamma^\prime = (\alpha - \beta) , \qquad 
(\alpha^\prime - \beta^\prime) = \gamma , \qquad 
\alpha^\prime/\beta^\prime = (1- \beta/\alpha)^{-1} , \qquad 
\tpar^\prime = - (\alpha/\beta - 1) \tpar \, , 
\label{eq:dualrep} 
\ee 
where the primes denote the complementary variables. The sign
transformation of the parametric time variable is necessary to make
the spirographic pattern turn in the counterclockwise sense; the
factor $(\alpha/\beta - 1)$ is not necessary but it keeps parametric
time passing at the same rate.  For completeness, we note that
epicycloids with $\beta < 0$ are also allowed and correspond to the
smaller circle rotating on the outside of the larger circle (although
such solutions are not considered here).

The parametric equations (\ref{eq:spirodef}) completely specify 
the geometry of the orbit (within the spirographic approximation). 
The velocity is given by the derivatives of the parametric equations 
with respect to physical time, i.e., 
$$ 
v_x = - \Bigl\{(\alpha - \beta) \sin \tpar + \gamma (\alpha/\beta - 1) 
\sin [(\alpha - \beta) \tpar /\beta] \Bigr\} {d \tpar \over dt } \, ,  
$$ 
\be 
v_y = \Bigl\{ - (\alpha - \beta) \cos \tpar + \gamma (\alpha/\beta - 1) 
\cos [(\alpha - \beta) \tpar /\beta] \Bigr\} {d \tpar \over dt } \, . 
\label{eq:spirospeed} 
\ee  
The direction of the velocity vector is thus determined (independent
of the derivative $d \tpar/dt$) for any point along the orbit. Since 
the radial displacement is given by $\xi = (x^2 + y^2)^{1/2}$, 
the magnitude of the velocity is determined from conservation of 
energy, which can be written in the dimensionless form 
\be 
v^2 = v_x^2 + v_y^2 = {1 \over 1 + \xi} - \epsilon \, , 
\ee 
where $v^2$(physical) = $v^2 (2 \Psi_0)$. 
We can complete the description by using this result to find the 
variation of parametric time $\tpar$ with physical time $t$, i.e., 
\be 
{d \tpar \over dt} = \Bigl[ {1 \over 1 + \xi} - \epsilon \Bigr]^{1/2}  
\Bigl[ \gamma^2 (\alpha/\beta - 1) \alpha/\beta + 
(\alpha - \beta)^2 \alpha/\beta - 
(\alpha/\beta - 1) \xi^2 \Bigr]^{-1/2} \, . 
\label{eq:dtdt} 
\ee 
This expression gives $d\tpar/dt$ in dimensionless form. This
formalism can be converted into physical units, where $d\tpar/dt$ has
units of inverse time, using the fiducial time scale defined by 
$t_0^2 \equiv (r_s/2 \Psi_0)$. 

The full orbits in this problem sweep through the angular coordinate
faster when the orbiting body is near the inner turning point (and
slower near the outer turning point). In this approximation scheme,
both the parametric expressions (eqs. [\ref{eq:spirodef}] and
[\ref{eq:spirospeed}]) and the conversion factor $d\tpar/dt$ have 
this qualitative behavior. In other words, some of the speed-up is
accounted for in the parametric equations and the rest is contained 
in $d\tpar/dt$. 

\subsection{Comparison with Physical Orbit Solutions}

One way to compare the dynamics of the physical problem and the
spirographic approximation is to write the effective equations of
motion in the same form. The equations of motion for the orbit problem
itself are given in the text above. The equation of motion for the
spirograph equivalent system can be written in the form 
\be 
{d \theta \over d \xi} = {1 \over \xi} 
{\xi_1 \xi_2 + (1- 2 \beta/\alpha) \xi^2 \over 
\Bigl[ (\xi^2 - \xi_1^2) \, (\xi_2^2 - \xi^2) \Bigr]^{1/2} } \, , 
\ee 
where $\xi = (x^2 + y^2)^{1/2}$ is the radial coordinate in the orbital
plane. Here we consider the system to be spirograph equivalent if
the orbits have the same turning points ($\xi_1,\xi_2$) and the same
turning angle ($\Delta \theta$, which sets $\beta/\alpha$).

Next we note that the equation of motion for both physical orbits
and the spirographic orbits can be written in the form 
\be
{d \theta \over d \xi} = {1 \over \xi} 
\Bigl[ (\xi - \xi_1) \, (\xi_2 - \xi) \Bigr]^{-1/2}  \, g(\xi) , 
\label{eq:general} 
\ee 
where the functions $g(\xi)$ are slowly varying over the radial range
of the orbits $\xi_1 \le \xi \le \xi_2$. We denote the functions
$g(\xi)$ as {\it distortion functions} because they determine the
manner in which the orbits are distorted from an elliptical shape.
For physical orbits in a Hernquist potential and the corresponding
spirographic approximation, the distortion functions $g(\xi)$ take 
the form 
\be 
g_{\rm phys} (\xi) = \Bigl[ {q \over \epsilon} 
{1 + \xi \over a + \xi} \Bigr]^{1/2} 
\qquad {\rm and} \qquad 
g_{\rm spi} (\xi) = {\xi_1 \xi_2 + (1 - 2 \beta/\alpha) \xi^2 \over 
\bigl[ (\xi + \xi_1) (\xi + \xi_2) \bigr]^{1/2} } \, . 
\label{eq:distortion} 
\ee
Notice that for Keplerian orbits, one obtains the same general form as
equation (\ref{eq:general}) with $g(\xi) = \sqrt{q/\epsilon} = 
\sqrt{ \xi_1 \xi_2}$ = {\sl constant}. The function $g_{\rm phys}$ 
reduces to this form in the limit $a \to 1$.  The corresponding 
function $g_{\rm spi}$ approaches the Keplerian form in the limit 
of low energy ($\epsilon \to 0$) and high angular momentum 
($q \to q_{max}$). 

Figures 5 and 6 show how closely the actual (physical) orbits can be
approximated by the spirographic treatment developed here.  Figure 5
shows the radial excursion of an orbit, with the dimensionless radius
plotted as a function of angle, for a physical orbit and the
corresponding spirographic orbit.  The two curves are nearly
indistinguishable, which vindicates the approximation for this system
($\epsilon$ = 0.90, $q/q_{max}$ = 0.75).

Next we need to quantify the difference between the physical orbits
and the spirographic approximation to the orbits over the entire 
parameter space. We can characterize the available parameter space in
terms of the quantities $(\epsilon, q/q_{max}$), where both variables
have the range $0 \le \epsilon, q/q_{max} \le 1$.  The difference
between the two orbits at a particular radius $\xi$ can be measured by
the quantity 
\be 
{\Delta g \over g} \equiv {g_{\rm phys} (\xi) - g_{\rm spi} 
(\xi) \over g_{\rm phys} (\xi) } \, , 
\label{eq:deltag} 
\ee 
where the distortion functions $g$ are defined above. The difference
between two orbits can be measured by the root mean square (RMS) of
the difference $(\Delta g)/g$ averaged over an entire orbit, i.e.,
over the range of radii $\xi_1 \le \xi \le \xi_2$. For a benchmark
comparison, we find the portion of parameter space for which the RMS
difference is less than 10\%. This value of 10\% is arbitrary. 
However, in many applications, such as the LMC orbit considered in the
following section, the observational errors are typically in the range
10 -- 20\%.  The result is shown in Figure 6. For the region in the
plane above the solid curve, the orbits can be represented as
epicycloids (spirograph patterns) with an effective error less than
10\%. Figure 6 also shows the portion of the $\epsilon-q$ plane for
which the orbits are close to Keplerian in the same sense.  The Kepler
approximation is also characterized by a function $g$ (see above), and
the dashed curve marks the locus of points for which the Keplerian
approximation differs from the physical orbit by 10\%. The Keplerian
approximation is thus valid only for the portion of parameter space to
the left and above the dashed curve. The bottom portion of the
$\epsilon-q$ plane shown in Figure 6 corresponds to low angular
momentum and hence nearly radial orbits.  The portion of parameter
space below the dot-dashed curve can be modeled, again within 10\%, as
radial orbits (for which we have analytic solutions -- see \S 3.4). 
Notice that for sufficiently deep orbits ($\epsilon \gta 0.9$), the
combination of the spirographic and radial approximations is
sufficient to describe the orbits; this portion of parameter space
corresponds to $\xistar \lta 0.073$. Finally, the location of the
orbit of the LMC is shown in the diagram, and is found to be squarely
within the spirographic regime (as discussed in \S 4).

Another way to assess the validity of the spirographic approximation
is to see how well the orbits conserve energy and angular momentum.
Conservation of energy is built into the approximation (through the
conversion factor $d\tpar/dt$) and is satisfied exactly. However, the
spirographic orbits follow slightly different paths (than the physical
orbits) and experience small variations in angular momentum over the
course of an orbital cycle. Within the spirographic formalism, the
angular momentum in dimensionless form is given by $j_{\rm spi}$ = $|
x v_y - y v_x | d\tpar/dt$, which can be written 
\be 
j_{\rm spi} = \Bigl[ {1 \over 1 + \xi} - \epsilon \Bigr]^{1/2} 
{ \gamma^2 (\alpha/\beta - 1) - (\alpha - \beta)^2 + 
\gamma (\alpha - \beta) (\alpha/\beta - 2) \cos (\alpha \tpar /\beta) 
\over \bigl[ \gamma^2 (\alpha/\beta - 1)^2 + (\alpha - \beta)^2 
- 2 \gamma (\alpha - \beta) (\alpha/\beta - 1) 
\cos (\alpha \tpar /\beta) \bigr]^{1/2} } \, .  
\ee
If the spirographic approximation were exact, this expression would be
constant. In general, the approximation scheme leads to a small
variation $(\Delta j)/j$ over the course of an orbit. The dotted curve
in Figure 6 shows the trajectory in parameter space for which the
angular momentum has an RMS deviation of 10\% over the course of
an orbit. In the region above the curve -- most of parameter space --
the variation in angular momentum is less than 10\%. Notice that
the requirement of angular momentum conservation, as enforced here, is
less stringent than that of orbital shape, as measured by the
distortion functions (compare the dotted and solid curves in Fig. 6). 
Since conservation of angular momentum is not precisely satisfied, the
spirographic formalism does not provide an exact solution to any
approximate physical problem; on the other hand, it does provide an
extremely good approximation to the exact physical problem.

\subsection{Orbital Decay under a Frictional Force} 

Within the spirographic approximation, we can calculate how orbits
change under the action of a frictional force.  As an example, we
consider the frictional force $\bf f$ per unit mass to be proportional
to the velocity so that 
\be
{\bf f} = - {1 \over T} {\bf v} \, ,
\ee
where the leading coefficient is constant and $T$ has units of time
(this form applies to the low speed limit of dynamical friction, e.g.,
see Binney \& Tremaine 1987).  The torque exerted on the orbiting body
is given by ${\bf r} \times {\bf f}$ and the time rate of change of
the specific angular momentum $j$ in terms of spirographic elements
becomes 
\be 
{d j \over dt} = - {r_s^2 \over T} \bigl\{ \eta \gamma^2 - 
(\alpha - \beta)^2 + (\alpha - \beta) \gamma (\eta - 1) 
\cos [ (1 + \eta) \tpar ] \bigr\} {d \tpar \over dt} \, , 
\ee 
where $\eta \equiv (\alpha - \beta)/\beta$ and where the conversion
factor $d\tpar/dt$ is given by equation (\ref{eq:dtdt}), including 
the fiducial time scale $t_0$. The change in angular momentum over 
each half orbit can be written  
\be 
\Delta j = \int_0^{\tau_{1/2}} dt \, {d j \over dt} = 
- {r_s^2 \over T} \int_0^{\beta \pi/\alpha} d\tpar \, 
\bigl\{ \eta \gamma^2 - (\alpha - \beta)^2 + 
(\alpha - \beta) \gamma (\eta - 1) \cos [ (1 + \eta) \tpar ] \bigr\} \, , 
\ee 
where we have changed the variable of integration (to parametric time) 
in the second equality. This procedure allows us to evaluate the 
integral to obtain 
\be 
\Delta j = - {r_s^2 \over T} \big[ \eta \gamma^2 - 
(\alpha - \beta)^2 \big] (\beta \pi / \alpha) \, . 
\label{eq:deltaj} 
\ee 
For comparison, in the limit of circular orbits (where $\alpha, \beta
\to 0$ and $\eta \to 1$) the change in angular momentum reduces to
$\Delta j = - (r_s^2/T) \gamma^2 (\pi/2)$.  This latter form is that
expected from applying a constant torque over a specified time (a half
orbital period). The difference between the circular expression and
equation (\ref{eq:deltaj}) shows how the geometry of the orbit affects
the decay of angular momentum. Within the spirographic approximation,
this expression for the change in angular momentum is exact.

Next we consider the loss of orbital energy. The work done on the 
orbiting body by the frictional force over a half orbit leads to a 
loss of energy, which can be written  
\be
\Delta E = \int {\bf f \cdot v} dt = - 
{r_s^2 \over T} \int_0^{\tau_{1/2}} dt \, 
\bigl\{ \gamma^2 \eta^2 + (\alpha - \beta)^2 - 
2 \gamma \eta (\alpha - \beta) \cos [ (1 + \eta) \tpar] \bigr\} 
\bigl( {d \tpar \over dt} \bigr)^2 \, . 
\ee
The integral can be simplified by changing the integration variable 
to parametric time. In this case, however, one factor of $d \tpar/dt$ is 
left over. Invoking the Mean Value Theorem, we can pull the extra factor 
out of the integral and the expression for $\Delta E$ becomes 
\be 
\Delta E = - {r_s^2 \over T} \langle {d \tpar \over dt} \rangle 
\bigl[ \gamma^2 \eta^2 + (\alpha - \beta)^2 \bigr]  (\beta \pi / \alpha)  \, . 
\ee
If we could identify the correct value of $\langle d\tpar / dt \rangle$, 
then the above expression would be exact within the spirographic formalism. 
In practice, we have to settle for an approximation; the mean value 
$\langle d\tpar / dt \rangle \approx (\beta \pi / \alpha \tau_{1/2})$ 
will suffice for most applications. In summary, the loss of energy per 
half orbit becomes 
\be 
\Delta E \approx - {r_s^2 \over T \, \, \tau_{1/2} } 
\bigl[ \gamma^2 \eta^2 + (\alpha - \beta)^2 \bigr]  
(\beta \pi / \alpha)^2  \, . 
\label{eq:deltae} 
\ee 
In addition to the approximation made in using the spirographic form
(which typically leads to errors of a few percent), this expression for 
energy loss contains a few percent error incurred in the evaluation of 
the integral. 

\section{THE ORBIT OF THE LARGE MAGELLANIC CLOUD} 

As an application of the solutions found here, we consider the orbit
of the LMC in the potential of the Galaxy. Using observational data in
conjunction with the treatment developed above, we find the orbital
elements of the LMC motion. The closest equivalent spirographic
approximation reproduces the physical orbit to a precision of $\sim
6-7\%$, and we thereby obtain a parametric (and hence analytic)
description of the orbit.

For this demonstration, we take the following observed quantities as
given: The observed velocities of the motion of the LMC in galactic
coordinates (van der Marel 2002) are as follows: radial velocity $v_r$
= 84 $\pm$ 7 km/s, tangential velocity $v_\theta$ = 281 $\pm$ 41 km/s,
and total velocity $v_T$ = 293 $\pm$ 39 km/s. The quoted errors lie in
the range $8-15$\% and result in uncertainties of comparable order in
the derived quantities of the orbit. The distance to the LMC is now
reasonably well known, with $\rlmc \approx$ 50 kpc (e.g., van der
Marel 2002).  We also need to specify the total mass $\massin$
contained within the LMC orbit; using velocity data from a large
number of Milky Way satellites, including the LMC, Kochanek (1996)
finds $\massin$ = $5 \pm 1 \times 10^{11} M_\odot$.  These quantities
are sufficient to specify the orbit within the framework developed
here, i.e., assuming that the potential of the Galaxy can be modeled
as a Hernquist potential over the range of radii probed by the LMC
orbit. For completeness, we note that in order to reproduce the proper
Galactic rotation curve at the solar circle, we would need to include
additional disk/bulge components which are neglected here (and which
introduce another source of uncertainty in the orbital parameters).

Conservation of energy implies 
\be 
{1 \over 2} v_T^2 = \Psi_0 \bigl[ - \epsilon + 
{1 \over 1 + \xilmc} \bigr] \, \approx (207 {\rm km/s})^2 \, ,  
\ee
where $\xilmc = \rlmc/r_s$. The scale $\Psi_0$ of the potential can be 
written in terms of known quantities, 
\be 
\Psi_0 = {G M_\infty \over r_s} = {G \massin \over \rlmc } 
{(1 + \xilmc)^2 \over \xilmc} \, \approx (207 {\rm km/s})^2 
{(1 + \xilmc)^2 \over \xilmc} \, . 
\ee
We can solve the above two equations for the dimensionless energy 
$\epsilon$ to obtain 
\be 
\epsilon = {1 + \xilmc - \xilmc/\mu \over (1 + \xilmc)^2} \, 
\approx {1 \over (1 + \xilmc)^2} \, , 
\label{eq:elmc} 
\ee 
where we have defined $\mu \equiv 2 G \massin / (\rlmc v_T^2)$. 
The current observational estimates suggest that $\mu \approx 1$, 
which leads to the second approximate equality. 
Using the definition of $q$, we can write 
\be 
q = \Bigl[ v_\theta^2 \bigl(2 G \massin / \rlmc \bigr)^{-1} \Bigr] 
{\xilmc^3 \over (1 + \xilmc)^2} \approx {0.92 \over \mu} 
{\xilmc^3 \over (1 + \xilmc)^2} \, , 
\label{eq:qlmc} 
\ee 
where we have used the observational results quoted above to 
obtain the numerical coefficient. 

If the scale length $r_s$ of the Galaxy and mass $\massin$ are known,
equations (\ref{eq:elmc}) and (\ref{eq:qlmc}) specify the energy and
angular momentum of the LMC orbit. The working estimate for $\massin$
uses data from many satellites of the Milky Way (Kochanek 1996) and is
reasonably secure (so that $\mu \approx 1 \pm 0.2$).  We estimate
$r_s$ as follows: Numerical simulations of the formation of dark
matter halos show that the density distributions approach a nearly
universal form (NFW) when properly scaled. A recent numerical study
(Navarro et al. 2004) indicates that the halo density profiles should
be scaled so that they line up at $r_{-2}$, the radius at which the
logarithmic slope of the density profile $s = d\log\rho/d\log r=-2$. 
The paper also shows that for halos with galactic masses (rotation
velocities $\sim200$ km/s), the radius $r_{-2} = 18 - 28 h^{-1}$ kpc.
For the Hernquist density profile used here, the logarithmic slope 
$s = -2$ occurs at $\xi = 1/2$ so that $r_s = 2 r_{-2}$ (note that
this relation is different for the original NFW profile where $r_{-2}
= r_s$). Using $h = 0.7$, we estimate the scale length of the Galaxy
to be $r_s \approx 65$ kpc.  As a result, $\xilmc = \rlmc/r_s \approx
0.77$ and our estimates for the energy and angular momentum of the LMC
orbit are 
\be 
\epsilon \approx 0.32 \qquad {\rm and} \qquad q \approx 0.134 \, \, 
(q/q_{max} \approx 0.69) \, . 
\ee 
The solid hexagon in Figure 6 marks the location of the LMC orbit in
the plane of possible orbits. Notice that the orbit falls within the
region where the spirographic approximation is valid. The observed
quantities used to estimate the orbital parameters are uncertain at
the $\sim10$\% level, so the resulting orbital elements are subject 
to comparable uncertainties.

Using the values of $\epsilon$ and $q$ found here, we can specify the
other parameters of the LMC orbit.  The inner turning point $r_1
\approx 46$ kpc, the outer turning point $r_2 \approx 114$ kpc, and
the effective semi-major axis of the orbit $r_\ast = r_s \xi_\ast
\approx 82$ kpc. With these estimates for the turning points, the
generalized eccentricities of the LMC orbit are $e = (\xi_2 -
\xi_1)/(2 \xi_\ast) \approx 0.41$ and $\evar = (\xi_2 - \xi_1)/(\xi_2
+ \xi_1) \approx 0.43$ The turning angle of the half orbit is given by
$\Delta \theta / \pi \approx 0.706$, so the orbit is far from closing.
The total radial period $\tau \approx 1.75$ Gyr, which is comparable
to the period of the equivalent circular orbit $\tau_c = 2 \pi
(\rlmc^3 / G \massin)^{1/2} \approx$ 1.5 Gyr.

For the estimates of $\epsilon$ and $q$ found here, the RMS deviation
of the physical orbit from its spirographic approximation is $\sim
6.6\%$. Because this difference is much smaller than the uncertainties 
in the orbit due to observational error and systematics, there is no
practical difference between the actual physical orbit and its
spirographic approximation.  For comparison, the RMS deviation of the
physical orbit from the Keplerian approximation is much larger (about
$17\%$). The spirographic orbital elements (in physical units) are
thus $\alpha$ = 48 kpc, $\beta$ = 14 kpc, and $\gamma$ = 80 kpc. The
corresponding spirographic orbital eccentricity $e_s$ = $(\alpha -
\beta)/\gamma$ $\approx$ 0.425.  Figure 7 shows an overlay of the
physical orbit and its approximate form given by the spirographic
formalism. The curves are shown in the orbital plane and the phases of
the orbits are aligned. The close agreement between the two orbits
argues that the spirographic approximation is adequate for modeling
the LMC orbit.

In the discussion thus far, we have assumed that the scale length
$r_s$ and mass $\massin$ of the Galaxy within the radius of the LMC
are known. The possible variation of these quantities provides an
important source of uncertainty in the orbital elements. For example,
if we assume that the scale length $r_s$ is completely unknown, then
equations (\ref{eq:elmc}) and (\ref{eq:qlmc}) constrain the LMC 
parameters to lie along the curve defined by  
\be 
q \approx 0.92 ( 1 - \sqrt{\epsilon} )^3 \, \epsilon^{-1/2}  \, . 
\ee 
If we allow the scale length $r_s$ to vary over the range $50 \le r_s
\le 80$, bracketing the estimated value, the allowed orbits in the
$\epsilon-q$ plane lie on the locus of points marked by open squares
in Figure 6.  Similarly, we can vary the mass of the Galaxy within the
current LMC position by varying the mass parameter $\mu$ in the above
equations $(\mu \equiv \massin /(5 \times 10^{11} M_\odot)$.  If we
let the enclosed mass vary over its allowed range $\mu = 1 \pm 0.2$ 
(Kochanek 1996), the orbit lies along the locus of points marked by 
open triangles in Figure 6.

This analysis implies a lower limit on the mass $\massin$. In order
for the LMC to be in a bound orbit, $\epsilon > 0$ and equation
(\ref{eq:elmc}) implies that 
\be 
\mu > {\xilmc \over (1 + \xilmc)} = {\rlmc \over \rlmc + r_s} \, . 
\label{eq:mulimit} 
\ee 
The scale length $r_s$ cannot be arbitrarily large. If we adopt a
conservative upper limit on $r_s$ based on the results of numerical
studies (e.g., Navarro et al. 2004, Bullock et al. 2001), then $r_s <$
100 kpc and hence $\mu > 1/3$. The corresponding mass limit becomes
$\massin > 1.7 \times 10^{11} M_\odot$. This limit is consistent with
previous estimates of Kochanek (1996) and others (van der Marel 2002;
Wilkinson \& Evans 1999; Lin, Jones, \& Klemola 1995).

Next we apply the frictional formulas developed in \S 3.3 to the orbit
of the LMC (see also Murai \& Fujimoto 1980).  The above analysis
determines the values for the spirographic parameters $\alpha$, $\beta$, 
and $\gamma$. Using these results in equation (\ref{eq:deltaj}), we 
find that the change in angular momentum over a half orbit is given by
$\Delta j \approx - 3.12 r_s^2/T$. The angular momentum itself is
given by $j = r_s \sqrt{2 q \Psi_0}$. Putting these results together 
with the parameters of the LMC orbit, we find that the fractional
change in angular momentum per orbit is given by 
\be
{\Delta j \over j} \approx - 0.94 \, \, 
\Bigl( {T \over 1 \, {\rm Gyr} } \Bigr)^{-1} \, . 
\ee 
The time scale $T$ can be evaluated using previously known results
(Binney \& Tremaine 1987). The new result found here is that the
change in angular momentum (per orbit) obtained using the true orbital
shape ($\Delta j \propto [\eta \gamma^2 - (\alpha - \beta)^2]
\beta/\alpha$) is larger than that of the equivalent circular orbit
($\Delta j \propto \gamma^2 / 2$) by 31\%.

\section{ORBITAL PROPERTIES FOR GENERAL MASS DISTRIBUTIONS}

The discussion thus far has focused on orbits in the Hernquist
potential and spirographic approximations to those orbits. In this
section we show that a wider class of potentials leads to orbits with
similar shape and can be adequately modeled using the spirographic
approximation. Toward this end, we prove a series of results that
apply to all physical potentials (\S 5.1 and 5.3) and find the region
of parameter space for which the spirographic approximation is valid
for a collection of particular potentials (\S 5.2).

\subsection{General Constraints on Orbital Shape} 

This section presents a set of general constraints on orbit shapes for
any extended mass distribution.  This set of constraints not only
delimits the allowed orbital paths, but also shows that spirographic
curves (epicycloids) provide reasonable approximations to the orbits
for general extended mass distributions. The potential for any such
mass distribution allows for two and only two turning points for bound
orbits (Contopoulos 1954). This constraint limits the orbital path to
lie within the annulus $\xin \le \xi \le \xout$. We also know that the
turning angle for a half orbit is confined to the range $\pi/2 \le
\Delta \theta \le \pi$ (Contopoulos 1954). To proceed further, we
assume that the turning points and the turning angle ($\xin$, $\xout$,
and $\Delta \theta$) are given, and find constraints on the path taken
as the orbit travels from ($\xout$, 0) to ($\xin$, $\Delta \theta$) in
the orbital plane. To illustrate these constraints, which are general,
we plot one particular example in Figure 8. Here, we use turning
points $\xin$ = 0.284, $\xout$ = 0.893, and a turning angle $\Delta
\theta$ = 2.01 (for a Hernquist profile, these values correspond to
the choices of dimensionless energy $\epsilon$ = 0.5 and angular
momentum $q/q_{max}$ = 0.5, but the constraints derived below hold for
any physical potential).

One can show that the orbits contain no inflection points (S. Tremaine, 
private communication). A related constraint can be found by using the 
definition of the turning angle and finding an upper bound on the
magnitude of the derivative $|d\xi/d\theta|$. Since the potential is
monotonic, $\psi \le \psi (\xin) = \epsilon + q/\xin^2$ over the radii
of interest, and this bound takes the form 
\be 
\Big| {d\xi \over d\theta} \Big|_{max} = \, 
\xi \bigl[ \xi^2 / \xin^2 - 1 \bigr]^{1/2} \, . 
\ee
The integral of this quantity produces a curve of minimum radius 
$\xi$ for a given angle $\theta$, and this curve has the form 
\be 
\xi (\theta) = \xin 
\csc \bigl[ \theta + \sin^{-1} (\xin/\xout) \bigr] \, .  
\ee
This function starts at the outer turning point and reaches the inner
turning point ``as soon as possible'', in the sense that any real
orbit must travel inward more slowly as a function of angle $\theta$.
This argument produces a complementary limiting form -- the orbit
could stay at the outer turning point for as long as possible and then
plunge inward (along a path of maximum $|d\xi/d\theta|$) just in time
to reach the inner turning point at $\theta = \Delta \theta$. The true
orbit is confined to lie between the two aforementioned curves, which
are shown as the solid lines in the illustrative example of Figure 8.

We can also consider the slowest possible turning of the orbit. The
turning angle $\Delta \theta \le \pi$ (Contopoulos 1954) and only
attains the value of $\pi$ in the Keplerian limit. The proof of this
result shows that the magnitude of the derivative $|d\xi/d\theta|$ is
always greater than that of the Keplerian orbit.  As a result, the
orbit must fall (from the outer turning point towards the inner
turning point) faster than the Keplerian orbit with the same turning
points, and the true orbit must lie within the Keplerian ellipse. As
before, this argument provides a complementary constraint: If we
construct a Keplerian orbit with the same turning points that reaches
the inner turning point at $\theta = \Delta \theta$, and trace the
orbit backwards out to the outer turning point, the physical orbit
must lie outside the curve. The physical orbit must thus lie between
these two ellipses, which are shown as dashed curves in Figure 8.

In general, as the density profile become less centrally concentrated,
the turning angle decreases. In particular, $\Delta \theta = \pi$ in
the limit of a point mass (the Keplerian limit), and the turning angle
$\Delta \theta = \pi/2$ for a uniform density distribution ($\rho$ =
{\sl constant}).  Since physical orbits are confined to have turning
angle $\Delta \theta \ge \pi/2$, we argue that the uniform density
orbit falls toward the inner turning point faster than any other case.
The uniform density model thus provides a benchmark for comparison and
its orbits are determined by the function $f_c (\xi)$ defined by 
\be 
f_c (\xi) = (\xi^2 - \xin^2) (\xout^2 - \xi^2) \, . 
\label{eq:fconst} 
\ee
We want to show that physical orbits are bounded by the orbit of the
uniform density model, with the turning points $\xin, \xout$ chosen to
coincide with those of the physical orbit.  Consider the difference in
angle between an arbitrary physical orbit [determined by $f(\xi)$] and
the constant density benchmark case [determined by $f_c (\xi)$].  We
define the difference in turning angle at a given radius to be $Q(\xi)
\equiv \theta (\xi; f) - \theta (\xi; f_c)$ so that  
\be 
Q(\xi) = \int_\xi^{\xout} {d\xi \over \xi} 
\Biggl( {\sqrt{q} \over f^{1/2}} - 
{\xin \xout \over f_c^{1/2}} \Biggr) \, .
\ee 
Now we look for the maximum angular difference. The derivative 
$dQ/d\xi$ = 0 if and only if $f(\xi)/q = f_c (\xi)/(\xin \xout)^2$ 
at some intermediate point $\xin < \xi < \xout$ (recall that $f/q=
f_c/(\xin\xout)^2$ at the turning points $\xin,\xout$ by construction). 
Next one can show that $f(\xi)/q = f_c(\xi)/(\xin\xout)^2$ at an
intermediate point only if the second derivative of the physical
potential $d^2 \psi/d\xi^2$ is less than that of the constant density
potential (since $d^2 \psi_c /d \xi^2 < 0$ for the constant density
case, ``less than'' means larger in magnitude and negative).  Finally,
we argue that the constant density potential has the greatest negative
curvature of any physically relevant potential, so that the derivative
$dQ/d\xi$ $\ne$ 0. This result implies that the difference in turning
angle between any physical orbit and the orbit of the constant density
potential must be monotonic. By inspection we find that $Q$ is
monotonically increasing.  As a result, physical orbits are confined
to lie outside the limiting curve found by integrating the orbit of a
constant density potential (see eq. [\ref{eq:fconst}]), i.e.,  
\be 
\theta(\xi) = {\pi \over 4} - {1\over2} \sin^{-1} \Bigl[ 
{ (\xin^2 + \xout^2) \xi^2 - 2 \xin^2 \xout^2 \over \xi^2 
(\xout^2 - \xin^2) } \Bigr] \, . 
\label{eq:rhocon} 
\ee
As before, a complementary bound exists. Since the orbit cannot turn
faster than equation (\ref{eq:rhocon}), one can construct a solution
in which the orbit stays at the outer turning point as long as
possible and then follows a function of the form (\ref{eq:rhocon}) to
the inner turning point. Physical orbits must lie within such a
curve. This argument leads to the pair of dotted curves shown in
Figure 8.

The above constraints limit the trajectory taken by an orbit in any
extended mass distribution. Each of the constraints leads to a pair of
bounding curves, with the true physical orbit confined to lie between
each pair of curves. The resulting combination of constraints implies
an orbit much like that of the epicycloid (spirographic) curve.  This
set of constraints is illustrated in Figure 8 for one particular case,
but this class of constraints holds for all orbits. 

\subsection{Allowed Parameter Space for Specific Potentials} 

Another way to illustrate the usefulness of the spirographic
approximation is to explicitly calculate the allowed regions of
parameter space for a collection of extended mass distributions. In
addition to the Hernquist profile discussed above, we calculate orbits
for three additional potentials and find the regimes of parameter space 
for which the spirographic approximation is valid. 
The first additional model is the NFW profile, where the density
distribution and potential can be written in the dimensionless form 
\be 
\rho = {1 \over \xi (1 + \xi)^2} \, \qquad {\rm and} \qquad 
\psi = {1 \over \xi} \ln (1 + \xi) \, . 
\ee
Next we consider a density distribution that is more concentrated 
and has the form 
\be 
\rho = {1 \over \xi^{3/2} (1 + \sqrt{\xi})^4 } \, \qquad 
{\rm and} \qquad \psi = {1 \over (1 + \sqrt{\xi})^2 } \, .
\label{eq:adams} 
\ee 
We denote this case as the 3/2 model (from the power-law slope of its
density profile in the inner limit).  Like the Hernquist model, these
potentials reach a finite central value and we have scaled the
dimensionless fields so that $\psi(0)$ = 1. As a result, the energy
$\epsilon$ is confined to the range $0 \le \epsilon \le 1$.  For a
given energy, each potential allows a maximum value $q_{max}$ of the
angular momentum parameter ($q_{max}$ corresponds to a circular orbit)
and the $q$ parameter is confined to the range $0 \le q/q_{max} \le 1$. 

We also consider the Jaffe model, where the density profile has the
form of a singular isothermal sphere for small radii, but falls more
quickly at large radii and reaches finite total mass (Jaffe 1983). 
These profiles have the form 
\be
\rho = {1 \over \xi^2 (1 + \xi)^2 } \, \qquad {\rm and} \qquad 
\psi = V_0 \ln \big( {1 + \xi \over \xi} \big) \, . 
\ee 
Unlike the previous cases, this potential does not reach a finite
value at the center and hence the dimensionless energy $\epsilon$ does
not have a finite range. The other potentials described above are
defined over the energy range $0 \le \epsilon \le 1$ and allow orbits
over the full radial extent $0 \le \xi \le \infty$.  In order to
compare the Jaffe model with these other potentials, we adopt the
value $V_0$ = 1/5; with this normalization, the orbits for $\epsilon$
= 1 have an effective semi-major axis $\xistar \approx$ 0.0041
(instead of $\xistar \to 0$).

For each potential defined above, we have calculated the orbits.  For
each physical orbit, we obtain the turning points ($\xi_1, \xi_2$) and
turning angle $\Delta \theta$, and use the results of \S 3 to define a
spirographic approximation to the orbital path. The departure of the
true orbit from its spirographic approximation is then calculated
using the method of distortion functions developed in \S 3.2.  For
these potentials, Figure 9 shows the region of parameter space for
which the spirographic approximation to the orbits is valid, with an
accuracy of (at least) 10\%. The allowed region of the $\epsilon-q$
plane lies above the various curves: Hernquist profile (solid curve),
NFW (dashed curve), the 3/2 profile (dotted curve), and the Jaffe
model (dot-dashed curve; keep in mind that the parameter space for the
Jaffe model extends beyond the portion of the $\epsilon-q$ plane shown
here). At low energies corresponding to large radius orbits, the
spirographic approximation has a limited range of validity for all of
the potentials. At higher energy, the regime of parameter space for
spirographic orbits depends on the degree of central concentration of
the density profile. The Hernquist and NFW profiles have the same
density dependence for small $\xi$ ($\rho \sim \xi^{-1}$) and similar 
ranges of validity. In general, the region of validity of the
spirographic approximation shrinks as the power-law index of the inner
density profile increases (compare the previous models with the 3/2
model and the Jaffe model).  This finding, that less centrally
concentrated mass distributions give rise to orbits that are more
spirographic in shape, can be understood from analytical
considerations: In the limit of a uniform density distribution, the
turning angle of the orbits has the value $\Delta \theta$ = $\pi/2$
and the corresponding distortion function is identical to that of the
spirographic approximation (see eq. [\ref{eq:distortion}] in the limit
$\beta \to \alpha/2$).  For completness, we note that this analysis
only considers the shape of the orbits. In order to provide a complete
description of the dynamics, one must include the transformation
between parametric time and physical time (the analog of eq. 
[\ref{eq:dtdt}]), and this transformation depends on the potential. 

Each potential also has a region of parameter space for which the
Keplerian approximation is valid (these regions are not delimited
here). The density profiles for the Hernquist and Jaffe models
approach the form $\rho \sim \xi^{-4}$ at large radii and hence reach
the Keplerian regime for sufficiently large $\xi$ (low energy
$\epsilon$). In contrast, the enclosed mass for the NFW profile
diverges logarithmically and hence the Keplerian approximation is
never applicable. The 3/2 model lies in between these two cases, with
a narrow regime of parameter space (at the low $\epsilon$ edge of the
plane) for which the Keplerian approximation is valid.

\subsection{Definitions of Orbital Eccentricity} 

The definition of orbital eccentricity necessarily contains an
ambiguity for extended mass distributions. Specifically, we show here
that the two alternate definitions of eccentricity $e = (\xout -
\xin)/2 \xistar$ and $\evar = (\xout - \xin)/(\xout + \xin)$ are 
equivalent if and only if the potential is Keplerian. The first
definition $e$ measures eccentricity relative to the circular orbit
given by $\xistar$ and has a maximum value $e_{max} < 1$. The second
definition $\evar$ is not centered on the circular orbit, but attains
the full range $0 \le \evar \le 1$.  The first half of the argument is
well known -- the two definitions of eccentricity are the same for a
Keplerian potential. 

To complete the argument, we suppose that the claim is true so that 
$e = \evar$. Then $\xin + \xout = 2 \xistar$, where $\xistar$ is the
orbital radius of the circular orbit (for a given $\epsilon$), and
$\xin$ and $\xout$ are the turning points. For a general potential
$\psi (\xi)$, the turning points are given by the zeroes of the
function $f(\xi)$, defined by
\be 
f(\xi) = - \epsilon \xi^2 + \xi^2 \psi(\xi) - q \, .  
\label{eq:fgeneral} 
\ee
The radius $\xistar$ of the circular orbit is given by the condition
$df/d\xi$ = 0. In order for $\xin + \xout = 2 \xistar$ to be valid for
all orbits, the function $f(\xi)$ must be symmetric with respect to
the point $\xi = \xistar$, i.e., $f(\xistar + x) = f(\xistar - x)$ for
all $x \in [0,\xistar]$. If the function $f(\xi)$ is symmetric, all
derivatives of odd order must vanish at $\xi = \xistar$. The first 
derivative $df/d\xi$ vanishes by definition. The third derivative 
must also vanish and is given by 
\be 
{d^3 f \over d \xi^3} = \xi^2 {d^3 \psi \over d \xi^3} + 
6 \xi {d^2 \psi \over d \xi^2} + 6 {d \psi \over d\xi} = 0 \, . 
\label{eq:diffeq} 
\ee 
This condition does not depend on $\epsilon$ or $q$, but only on the
form of the potential. Furthermore, the possible range of $\epsilon$
and $q$ allow for all values of $\xistar$ to be realized.  In order
for the third derivative of $f(\xi)$ to vanish for all orbits,
equation (\ref{eq:diffeq}) must hold for all values of $\xi$.  This
differential equation is second order in the function $U \equiv
d\psi/d\xi$ and has two linearly independent solutions: $U_A =
1/\xi^2$ and $U_B = 1/\xi^3$.  The first solution corresponds to
$\psi_A \sim 1/\xi$ which is the Keplerian case. The second solution
leads to $\psi_B \sim 1 / \xi^2$, which implies a negative mass
density, and is thus unphysical. The original differential equation
(\ref{eq:diffeq}) is third order and has a third solution, $\psi_C$ =
{\sl constant}, which is not of interest. Thus, we have shown that if
$e = \evar$, then the potential $\psi = 1/\xi$ and is Keplerian.  In
order words, the validity of the condition $\xin +
\xout = 2 \xistar$ requires a Keplerian potential.

\section{CONCLUSION}  

This paper has explored orbital trajectories for the Hernquist
potential and other extended mass distributions. Our results can be
summarized as follows:

[1] The first set of results quantitatively determines the orbits for
the Hernquist potential. The analysis is straightforward.  The orbits
have a rosette shape (Fig. 1), as was well-known previously in
qualitative terms (Binney \& Tremaine 1987). The resulting orbits can
be described by their turning angles $\Delta \theta$ and half periods
$\tau_{1/2}$ for a given value of energy $\epsilon$ and angular
momentum $q$ (see Figs. 2 and 3). In analogy to the Kepler problem, we
have defined osculating orbital elements (Table 1). The Hernquist 
profile allows for a number of results to be derived analytically. 
We have found the energy and angular momentum as function of the
turning points (eq. [\ref{eq:inverseroots}]), the maximum angular
momentum for a given energy (eq. [\ref{eq:qmax}]), a radial scale
$\xistar$ that plays the role of the semi-major axis
(eq. [\ref{eq:xistar}]), the analog of Kepler's law (\S 2.2), an
analytic form for the turning angle in the limit of circular orbits
(eq. [\ref{eq:thetalim}]), and an expression for the radial period in
the limit of low angular momentum (eq. [\ref{eq:analtau}]). This
latter expression provides a good approximation to the radial period
for all orbits (Fig. 3). We have also constructed a fitting formula
(eq. [\ref{eq:fitting}]) that specifies the turning angle $\Delta
\theta (\epsilon, q)$ as a function of energy and angular momentum
over most of parameter space.

[2] This paper shows that the physical orbits of the Hernquist profile
can be modeled to a good approximation by epicycloid curves, more
commonly known as spirograph patterns. These curves have analytic
solutions, which can be written in the parametric form of equation
(\ref{eq:spirodef}). These parametric equations allow for a completely
analytic description of the orbits, and we have developed this
approximation in some detail. The transformation between spirographic
orbital elements and physical parameters is given by equations
(\ref{eq:relate}--\ref{eq:eqspiro}). The effective eccentricity of a
spirographic curve is given by equation (\ref{eq:eccentspiro}). Most
importantly, the spirographic approximation faithfully reproduces the
shape of the physical orbits to an accuracy better than 10\% over most
of parameter space (Fig. 6). The spirographic approximation conserves
energy exactly (using the transformation of eq. [\ref{eq:dtdt}] to
relate physical time to parametric time) and conserves angular
momentum to a greater accuracy than it reproduces the orbital shape.

[3] As a demonstration of the efficacy of this approach, we have found
the shape of the orbit of the Large Magellanic Cloud as it traces
through the halo of our Galaxy. If we model the Galactic potential
with a Hernquist potential, the dimensionless energy and angular
momentum of the LMC orbit are ($\epsilon, q/q_{max}$) = (0.32, 0.69)
and the orbit takes the form shown in Figure 7. The spirographic
approximation conserves angular momentum to an accuracy of 1\%
and reproduces the shape of the physical orbit to an accuracy of 6 --
7\%. These levels of error are much smaller than both the
observational uncertainties in the problem and the errors incurred in
approximating the Galaxy as a smooth, spherical, extended mass
distribution. As a result, the analytic (spirographic) approximation
provides a good working model for the orbit (with essentially no loss
of accuracy).

[4] In addition to considering orbits in the Hernquist potential, this
paper argues that the results found here -- in particular the
spirographic approximation to the orbits -- apply to general extended
mass distributions. Toward this end, we have presented general
constraints on the orbital shape (\S 5). Figure 8 shows that the
orbits are restricted to be relatively close to the shape of an
epicycloid for any extended mass distribution. Figure 9 shows that
the spirographic approximation is valid over much of the $\epsilon-q$
plane for a collection of specific potentials (including the NFW
profile). We have also shown that the definition of orbital
eccentricity necessarily contains an ambiguity: For a given mass
distribution, the eccentricity can either be defined to be symmetric
with respect to a radial scale that plays the role of the semi-major
axis, or the it can be defined solely in terms of the turning points
(see eq. [\ref{eq:eccent}]); the two definitions coincide only for a
Keplerian potential (\S 5.3).

Many astrophysical systems involve orbits in extended mass
distributions and an analytic approximation to the orbits is often a
useful tool. This paper develops the spirographic approximation, which
can be used to model orbits in a wide variety of contexts. One such
example, that of the LMC orbit, has been studied in this paper, but
many additional applications of this formalism remain. Possible future
topics include orbits of galaxies within their clusters, orbits of
dark matter particles for dark matter detection strategies, globular
cluster orbits in galactic potentials, orbits of small halos as they
merge into larger halos during structure formation, orbits of young 
stars in their birth aggregates, and many others.

In addition its direct applications, the approach developed here can
be generalized in a number of ways. This work has focused on the
Hernquist potential, although some comparison with other models has
been included (see Fig. 9). Other cases should be considered and
compared with the spirographic approximation, including the HH model
in which two different Hernquist potentials are nested together
(Ciotti 1996), models with varying degrees of central concentration
(Tremaine et al. 1994), and the inclusion of a point mass at the
origin (to model the central black holes found in many galactic
centers). For power-law potentials, the turning angles have already
been computed (Touma \& Tremaine 1997) and the spirographic
approximation can be readily implemented. The orbits in this paper are
confined to a single plane, for spherical potentials, but the
spirographic approximation can be implemented for two dimensional
orbits in nonspherical potentials.

We note that the Hernquist potential and alternate forms (e.g., the
NFW profile or the Jaffe model) are just models for more complicated
physical structures. In the case of galactic halos, for example, the
true potential must have corrections due to clumpiness and hence small
scale inhomogeneities; a galactic disk (or multiple disk components)
and hence a large scale quadrapole moment; perturbations from nearby
large galaxies (e.g., Andromeda); and other complications. Many of
these effects enter at the $\sim$10\% level and imply 10\% corrections
to the orbital shape calculated from a spherical potential. This paper
shows that over most of the relevant parameter space, the spirographic
approximation leads to orbital shapes that agree with those of the
Hernquist profile (and others -- see Fig. 9) to an accuracy of better
than 10\%. Thus, in practice, the spirographic approach can be used as
an approximation to the orbits without losing accuracy (compared to
numerically calculated orbits) while providing an analytic description
of the dynamics.

%\bigskip 
%$\,$ 
\newpage 

\centerline{\bf Acknowledgment} 

We would like to thank Dan Amidei, Michael Busha, Gus Evrard, Lars
Hernquist, and Scott Tremaine for beneficial discussions and other
useful input. We thank the referee for many useful comments that
improved the manuscript. This work was supported at the University of
Michigan by the Michigan Center for Theoretical Physics; by NASA
through the Terrestrial Planet Finder Mission (NNG04G190G) and the
Astrophysics Theory Program (NNG04GK56G0); and by NSF through grants
DMS01038545 and DMS0305837.

\newpage 
%\bigskip 
%$\,$  

\centerline{\bf REFERENCES} 

\par\pp 
Abramowitz, M., \& Stegun, I. A. 1972, Handbook of Mathematical
Functions (New York: Dover)

\par\pp
Adams, F. C., \& Laughlin, G. 2001, Icarus, 150, 151 

\par\pp
Adams, F. C., \& Myers, P. C. 2001, ApJ, 553, 744

\par\pp
Beyer, W. H. 1980, editor, Standard Mathematical Tables 
(Boca Raton: CRC Press) 

\par\pp
Binney, J., \& Tremaine, S. 1987, Galactic Dynamics 
(Princeton: Princeton Univ. Press) 

\par\pp 
Bullock, J. S., Kolatt, T. S., Sigad, Y., Somerville, R. S., Kravtsov,
A. V., Klypin, A. A., Primack, J. R., \& Dekel, A.  2001, MNRAS, 321, 559 

\par\pp
Busha, M. T., Adams, F. C., Wechsler, R. H., \& Evrard, A. E. 2003,
ApJ, 596, 713   

\par\pp
Busha, M. T., Evrard, A. E., Adams, F. C., \& Wechsler, R. H. 2004, 
submitted to MNRAS 

\par\pp
Ciotti, L. 1996, ApJ, 471, 68 

\par\pp
Contopoulos, G. 1954, Zeitschrift Astrophys, 35, 67 

\par\pp 
Crone, M., Evrard, A. E., \& Richstone, D. O. 1996, ApJ, 467, 489 

\par\pp
Dehnen, W. 1999, AJ, 118, 1190  

\par\pp
de Vaucouleurs, G. 1948, Ann. d'Ap, 11, 247  

\par\pp
Evans, N. W., \& Read, J.C.A. 1998, MNRAS, 300, 83  

%\par\pp
%H{\'e}non, M. 1959, Ann. d'Astrophys, 22, 126 

\par\pp
Hernquist, L. 1990, ApJ, 356, 359     

\par\pp
Jaffe, W. 1983, MNRAS, 202, 995 

\par\pp
Jijina, J., Myers, P. C., \& Adams, F. C. 1999, ApJ Suppl., 125, 161  

\par\pp
Kalnajs, A. J. 1979, AJ, 84, 1697 

\par\pp 
Kochanek, C. S. 1996, ApJ, 457, 228 

\par\pp
Lada, C. J., \& Lada, E. A. 2003, ARA\&A, 41, 57 

\par\pp
Larson, R. B. 1985, MNRAS, 214, 379  

\par\pp
Lin, D.N.C., Jones, B. F., \& Klemola, A. R. 1995, ApJ, 439, 652 

\par\pp 
Moore, B., Governato, F., Quinn, T., Stadel, J., \& Lake, G. 1998,
ApJ, 499, L5 

\par\pp
Murai, T., \& Fujimoto, M. 1980, PASJ, 32, 581  

\par\pp
Murray, C. D., \& Dermott, S. F. 1999, Solar System Dynamics 
(Cambridge: Cambridge Univ. Press) 

\par\pp 
Navarro, J. F., Frenk, C. S., \& White S.D.M. 1997, ApJ, 490, 493 (NFW) 

\par\pp
Navarro, J. F., Hayashi, E., Power, C., Jenkins, A. R., Frenk, C. S.,
White, S. D. M., Springel, V., Stadel, J., \& Quinn, T. R. 2004, 
MNRAS, 349, 1039

\par\pp
Newton, I. 1687, Philosophia Naturalis Principia Mathematica 

\par\pp 
Porras, A., Christopher, M., Allen, L., DiFrancesco, J., Megeath, 
S. T., \& Myers, P. C. 2003, AJ, 126, 1916  

\par\pp 
Shu, F. H. 1969, ApJ, 158, 505 

\par\pp
Smale, S. 1970a, Inventiones Mathematicae, 10, 305 

\par\pp
Smale, S. 1970b, Inventiones Mathematicae, 11, 45 

\par\pp
Touma, J., \& Tremaine, S. 1997, MNRAS, 292, 905 

\par\pp  
Tremaine, S. et al. 1994, AJ, 107, 634 

\par\pp
van den Bosch, F. C., Lewis, G. F., Lake, G., \& Stadel, J. 1999, 
ApJ, 515, 50 

\par\pp 
van der Marel, R. P., Alves, D. R., Hardy, E., \& Suntzeff, N. B. 
2002, AJ, 124, 2369 

\par\pp
Wilkinson, M. I., \& Evans, N. W. 1999, MNRAS, 310, 645 

\newpage 
\begin{figure}
\figurenum{1}
{\centerline{\epsscale{0.90} \plotone{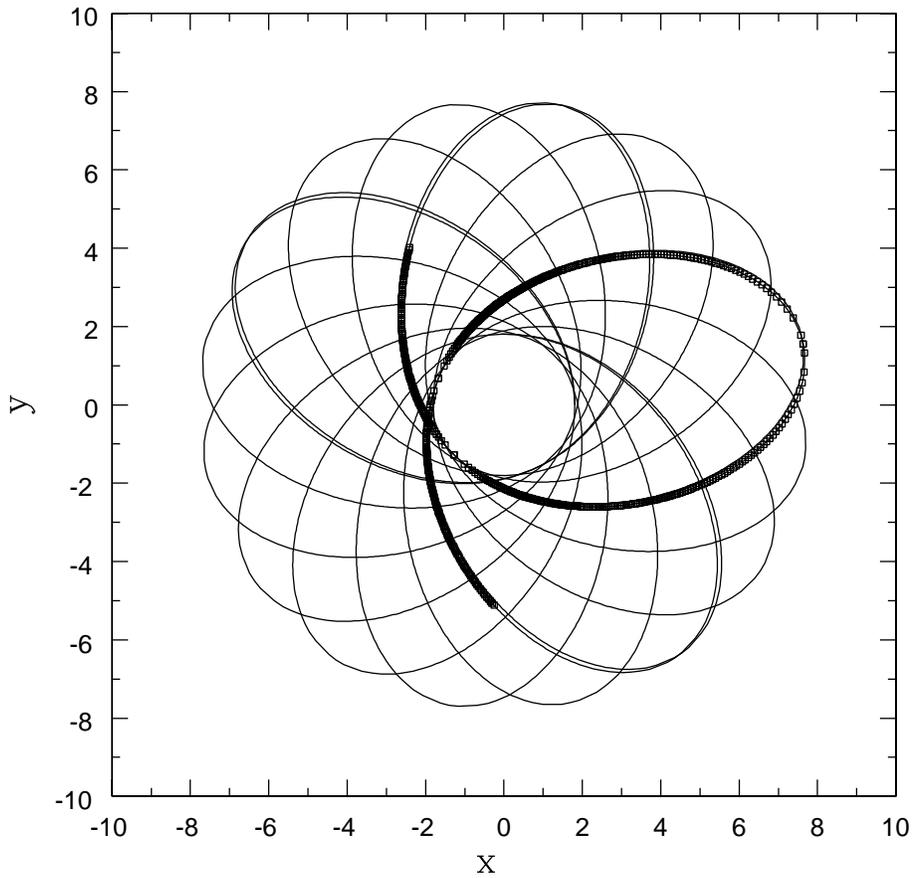} }}
\figcaption{An example of an orbit in the Hernquist potential with
energy $\epsilon$ = 0.10 and angular momentum parameter $q$ =
$q_{max}/2$. The orbit does not close and can be characterized by the
angle through which the orbit turns during one radial oscillation. One
part of the orbit is highlighted (with the open square symbols). }
\end{figure}

\newpage 
\begin{figure}
\figurenum{2}
{\centerline{\epsscale{0.90} \plotone{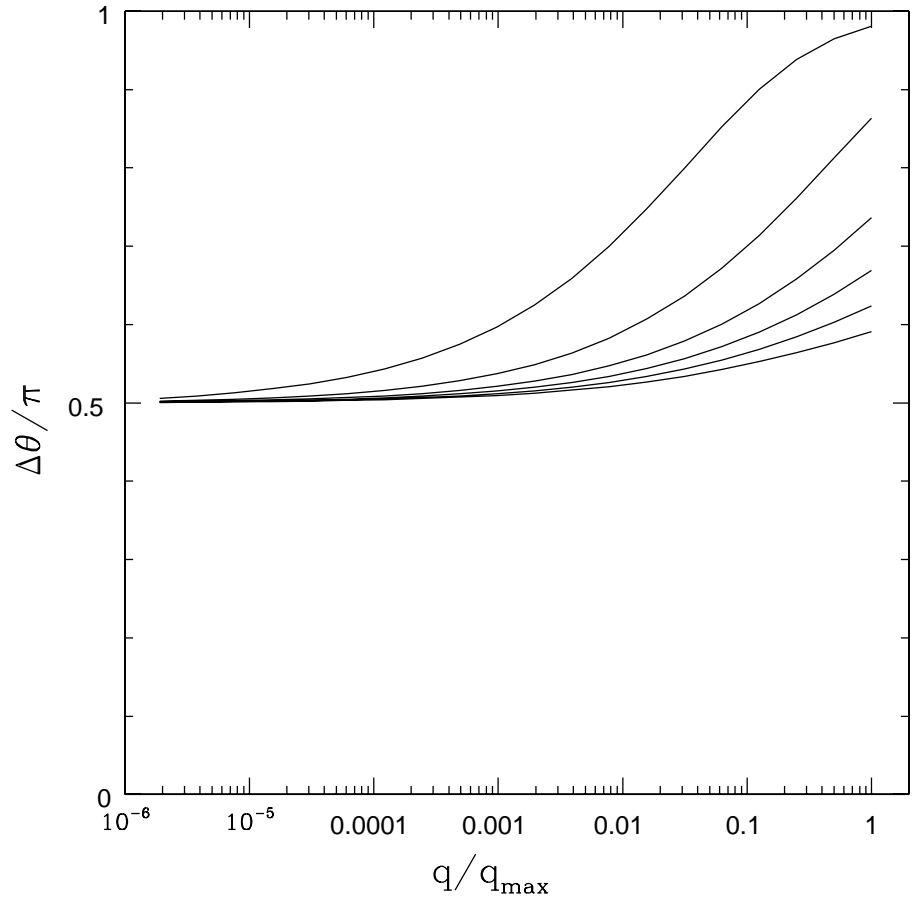} }}
\figcaption{The turning angles for one half of an orbit. The quantity
$\Delta \theta$ is the angle through which the orbit turns as the
radius varies from the inner turning point to the outer turning point.
The turning angle $\Delta \theta$ for a half orbit is plotted here as 
a function of $q/q_{max}$, where $q_{max}$ is the maximum angular
momentum accessible for a bound orbit at a given energy. The various
curves correspond to dimensionless energy values $\epsilon$ = 0.01,
0.1, 0.3, 0.5, 0.7, and 0.9 (from top to bottom in the figure). Notice
that $\Delta \theta \to \pi/2$ in the limit $q \to 0$, but, in
general, $\Delta \theta \ne \pi$ in the limit of circular orbits ($q
\to q_{max}$). }  \end{figure}

\newpage 
\begin{figure}
\figurenum{3}
{\centerline{\epsscale{0.90} \plotone{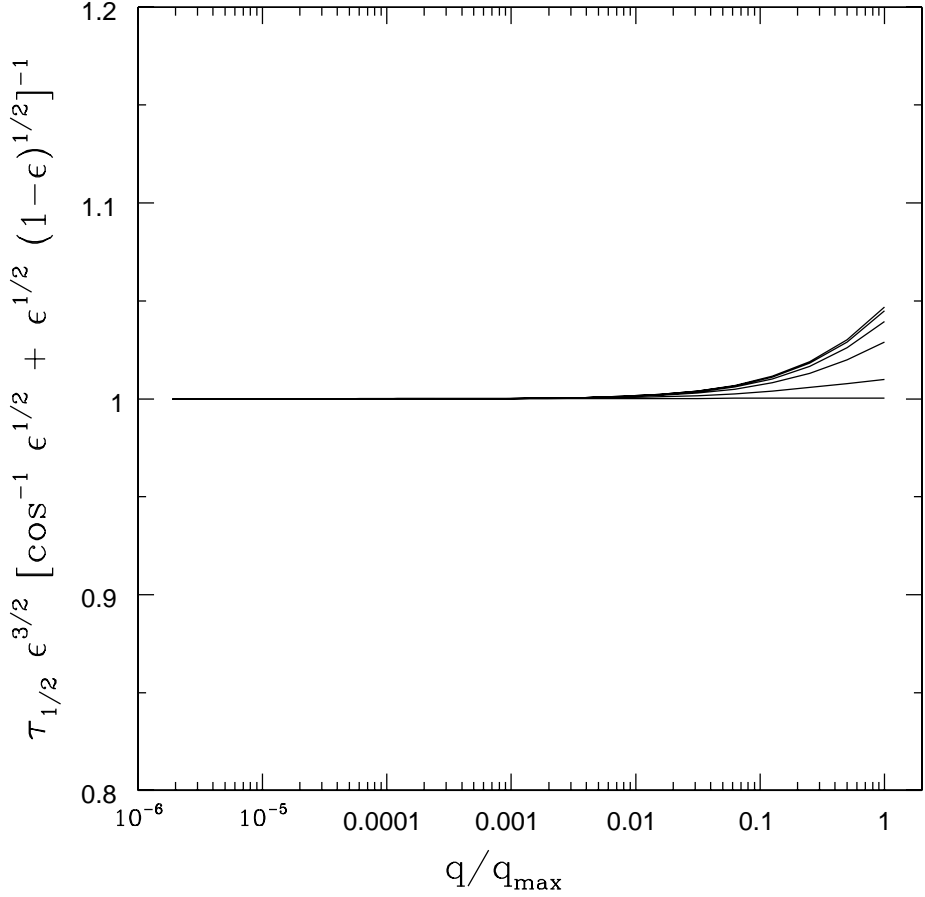} }}
\figcaption{The range of orbital half period. The quantity $\thalf$ is
the time required for the orbit to travel from the inner turning point
to the outer turning point. Here, the half period is normalized by two
functions, first the factor $\epsilon^{3/2}$, which represents the
variation expected in the Keplerian limit, and by the factor
$[\cos^{-1} \sqrt{\epsilon} - \sqrt{\epsilon} (1 - \epsilon)^{1/2} ]$,
which gives the correct orbit time in the limit of zero angular
momentum ($q \to 0$). The half period is plotted as a function of
$q/q_{max}$, where $q_{max}$ is the maximum angular momentum
accessible for a bound orbit at a given energy. The various curves
correspond to dimensionless energy values $\epsilon$ = 0.01, 0.1, 0.3,
0.5, 0.7, and 0.9 (from bottom to top in the figure). }  
\end{figure}

\newpage 
\begin{figure} 
\figurenum{4} 
{\centerline{\epsscale{0.90} \plotone{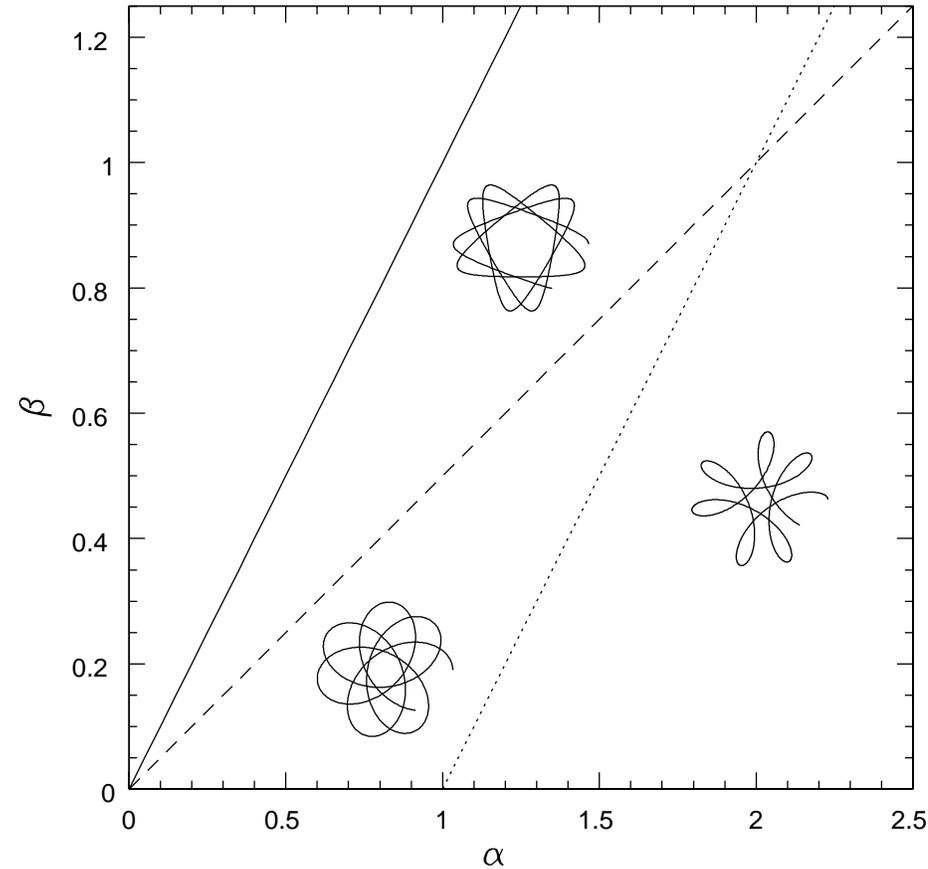} }}
\figcaption{Allowed range of parameter space for spirographic orbits
(epicycloids) to approximate physical orbits.  The drawing point
parameter $\gamma$, which plays the role of the semi-major axis, has
been set equal to unity. The parameter $\alpha$ represents the radius
of the larger circle and $\beta$ is the radius of the smaller circle.
The value of $\beta$ must lie below the dashed line (where $\beta =
\alpha/2$) so that the turning angle of a half orbit is greater than
$\pi/2$. The $\beta$ value must also lie above the dotted line (where
$\beta = \alpha - \gamma$) so that the inner turning point is
positive. Physical orbits are confined to the region below the dashed
line and above the dotted line. The inset diagrams show representative 
orbital shapes for the delimited regions of parameter space. The upper 
left portion of the diagram, above the solid line where $\alpha = \beta$, 
is not allowed ($\beta < \alpha$ by definition). The upper right portion
of the diagram (above the dashed line and below the dotted line) allows 
for an alternate representation of the physical orbits (see text). } 
\end{figure}

\newpage 
\begin{figure} 
\figurenum{5} 
{\centerline{\epsscale{0.90} \plotone{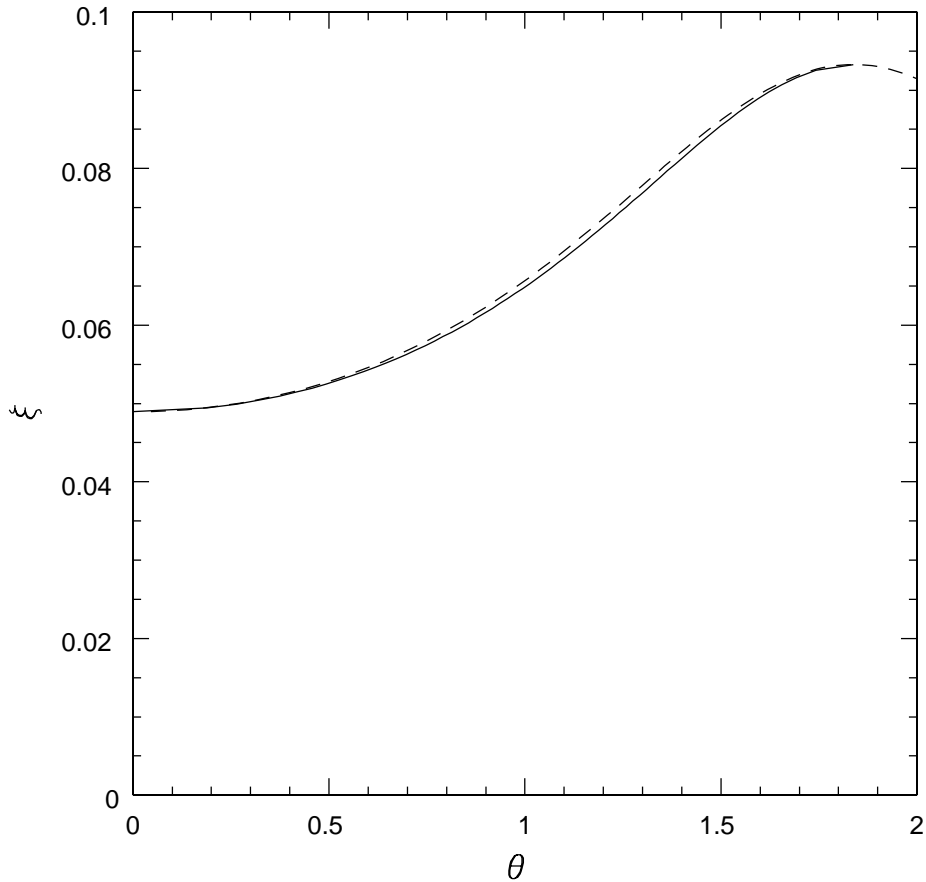} }}
\figcaption{Comparison of the radial coordinate as a function of turning
angle for the physical orbit problem in a Hernquist potential (solid
curve) and the equivalent spirograph orbit (dashed curve). The two
systems are chosen to have the same radial turning points ($\xi_1$ and
$\xi_2$) and the same angular displacement per half orbit $\Delta
\theta \approx 1.84$ ($\Delta \theta / \pi \approx 0.585$). These
values correspond to the energy $\epsilon$ = 0.90 and the angular
momentum variable $q/q_{max}$ = 0.75 for the Hernquist potential. 
According to the method of distortion functions developed in the text, 
the two orbit shapes differ by 2.2\%; the spirographic orbit conserves 
angular momentum to an accuracy of 0.3\%. For comparison, the shape 
of the physical orbit differs from that of a Keplerian orbit with the 
same turning points by 69\%. }  
\end{figure}

\newpage 
\begin{figure} 
\figurenum{6} 
{\centerline{\epsscale{0.90} \plotone{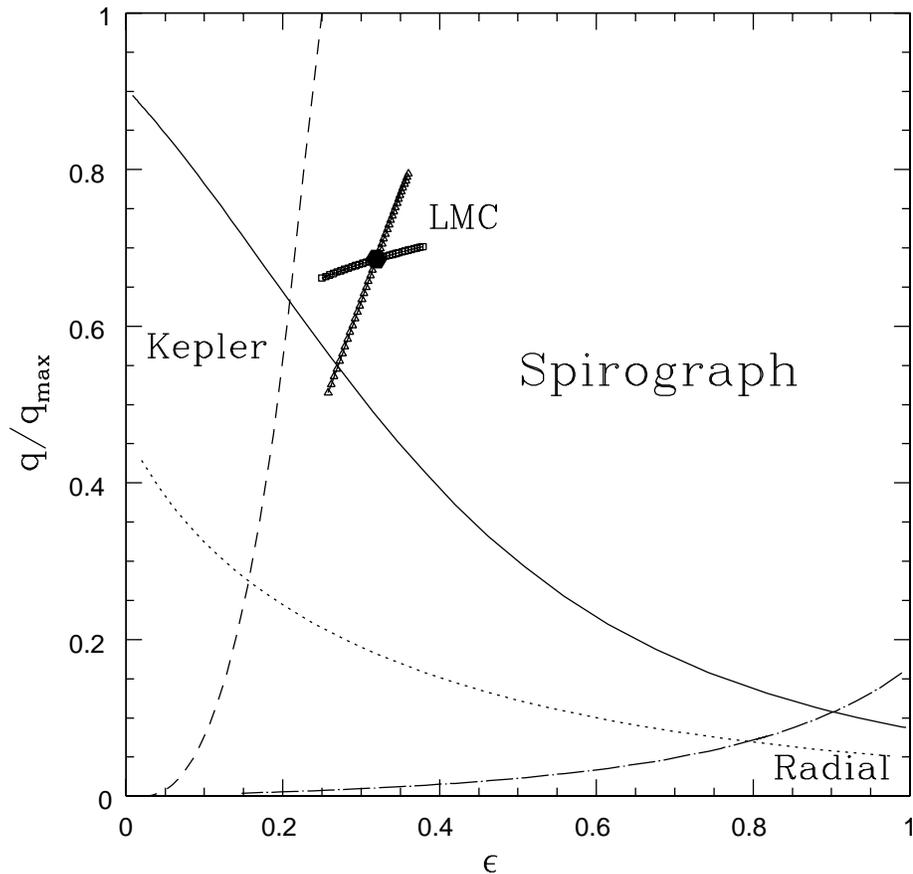} }}
\figcaption{Regions of parameter space for which the Keplerian and
spirographic approximations are valid for the Hernquist profile. 
The region of the $\epsilon-q$ plane above the solid curve corresponds
to orbital parameters for which the RMS deviation in the shape of the
physical orbits from that of the spirographic approximation to the
orbits is less than 10\%. Similarly, the region above the dashed
curve is where the RMS deviation of the physical orbits from Keplerian
orbits is less than 10\%. The region above the dotted curve is
where the spirographic approximation conserves the angular momentum of
the orbit to an accuracy of 10\%. Finally, the region below the
dot-dashed curve is where the orbits are radial to within 10\%. 
The large hexagon at $(\epsilon,q/q_{max}) \approx (0.32,0.69)$ 
marks our best estimate for the orbital parameters of the LMC. The 
accompanying (short) curves show the variation of the orbital 
parameters with variations in the Galactic scale radius $r_s$ 
(open squares) and enclosed mass $M_{\rm in}$ (open triangles). }
\end{figure} 

\newpage 
\begin{figure}
\figurenum{7}
{\centerline{\epsscale{0.90} \plotone{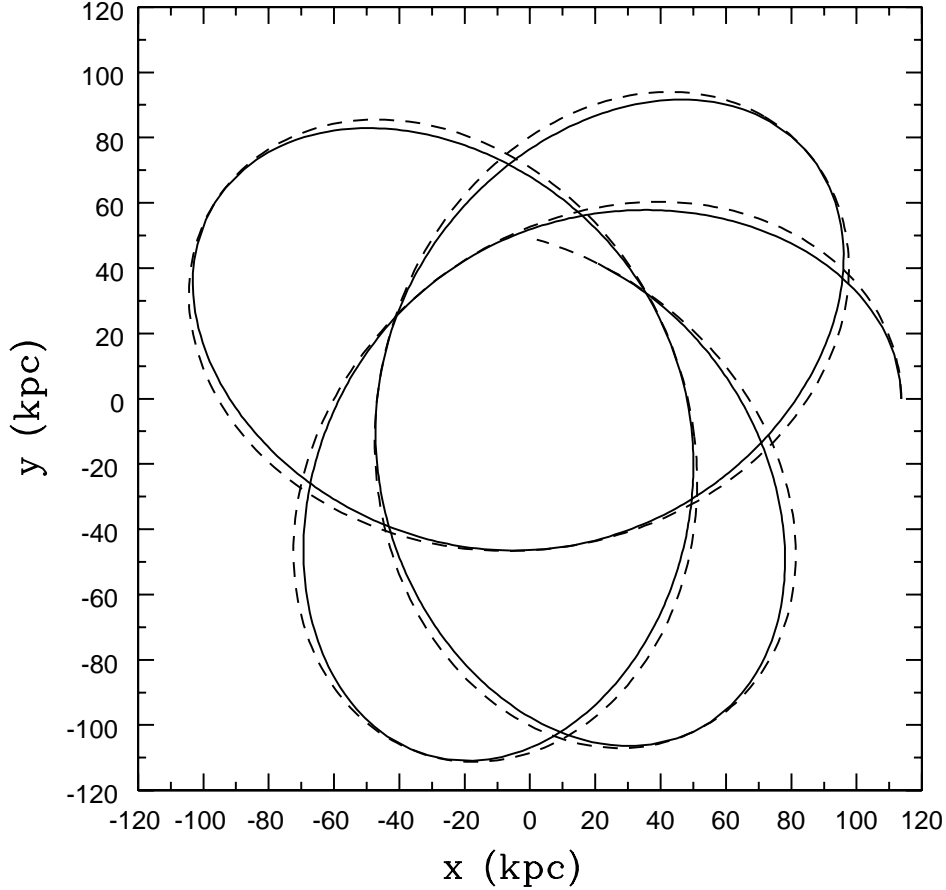} }}
\figcaption{The predicted shape of the LMC orbit in the potential of
the Galaxy. The coordinates $x$ and $y$ represent the plane of the
orbit and are expressed in kpc. The solid curve shows the orbital
shape for a Hernquist potential using our best estimate for the
dimensionless orbital energy and angular momentum $(\epsilon,
q/q_{max})$ = (0.32,0.69).  The dashed curves shows the spirographic
(analytic) approximation to the orbit. The shape of the orbits agree
to within about 7\%, as measured by the distortion functions
(eqs. [\ref{eq:distortion}, \ref{eq:deltag}]). As another measure, the
analytic (spirographic) orbit conserves angular momentum to an accuracy 
of 1\%. For comparison, the observational determination of the 
orbital parameters is subject to 10 -- 20\% uncertainties. }  
\end{figure}

\newpage 
\begin{figure}
\figurenum{8}
{\centerline{\epsscale{0.90} \plotone{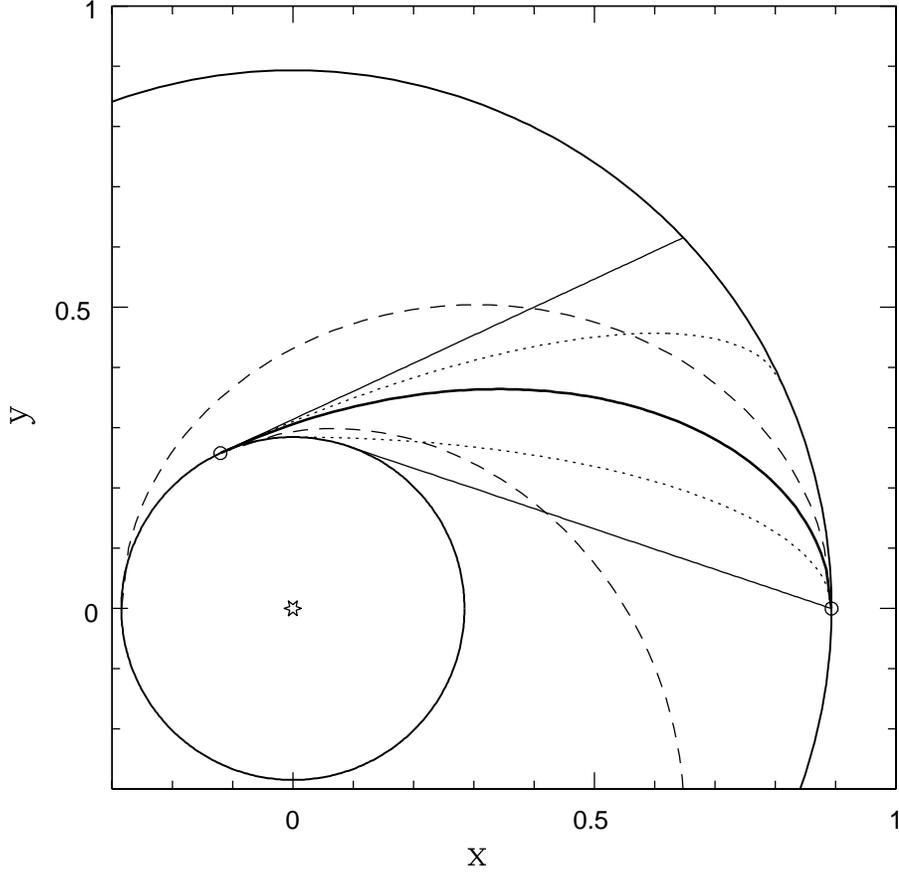} }}
\figcaption{General constraints on orbital shape for an arbitrary
potential. The bold-faced curve shows the spirographic orbit for given
turning points (marked by the open symbols, with radii marked by the
inner and outer circles) and a given turning angle for a half orbit.
The two solid lines represent an upper limit on how fast the orbit can
turn, so that the orbit must trace a path between the two lines. 
Similarly, the dashed curves depict a limit on how slowly the orbit
can turn; this limit corresponds to elliptical orbits and physical
orbits must fall between these curves as well. The dotted curves
represent the constraint that the orbit cannot turn any faster than
the orbit of the uniform density model (which has $\Delta \theta$ = 
$\pi/2$ and all orbits must have $\Delta \theta \ge \pi/2$). 
Physical orbits (for given $\xin, \xout, \Delta \theta$) must lie
between the three pairs of constraints, thereby limiting the orbital
shape to be reasonably close to that of the spirographic orbit. } 
\end{figure}

\newpage 
\begin{figure}
\figurenum{9}
{\centerline{\epsscale{0.90} \plotone{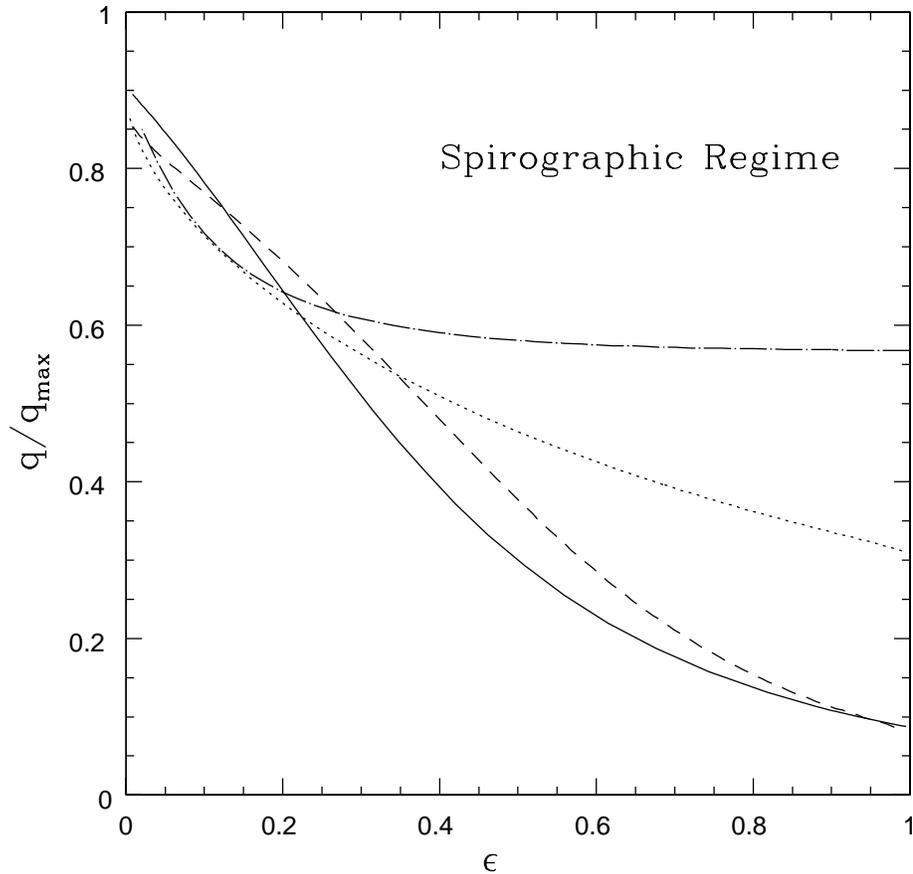} }}
\figcaption{ Regions of parameter space for which the spirographic
approximation is valid for a collection of potentials. The region of
the $\epsilon-q$ plane above the curves corresponds to orbital
parameters for which the shape of the physical orbits deviates from
that of the spirographic approximation by less than 10\%. The
potentials considered here are the Hernquist potential (solid curve),
the NFW profile (dashed curve), the 3/2 model with density profile
$\rho \sim \xi^{-3/2} (1 + \sqrt{\xi})^{-4}$ (dotted curve), and the
Jaffe model (dot-dashed curve). The Jaffe model differs from the other
cases in that the potential does not reach a finite central value and
hence the plane depicted here does not represent all of its parameter
space.}
\end{figure} 

\end{document}